\documentclass[aps,manuscript,overcite]{revtex}

\usepackage{graphics}
\input epsf

\def\a{a}
\def\n{n}
\def\H{H}
\def\D{D}
\def\m{m}
\def\t{\tau}
\begin{document}

\title{ The Ekpyrotic Universe:
Colliding Branes and the Origin of the Hot Big Bang }

\author{ Justin Khoury$^1$, Burt A. Ovrut$^2$, Paul J. Steinhardt$^1$ 
and Neil Turok $^{3}$}

\address{
$^1$ Joseph Henry Laboratories,
Princeton University,
Princeton, NJ 08544, USA \\
$^2$ Department of Physics, University of Pennsylvania,
Philadelphia, PA 19104-6396, USA\\
$^3$ DAMTP, CMS, Wilberforce Road, Cambridge, CB3 0WA, UK}

\maketitle

\begin{abstract}
We propose a cosmological scenario in which the hot big bang universe
is produced by the collision of a brane in  the bulk space
with a bounding orbifold
plane, beginning from an otherwise cold, vacuous, static universe.
The model addresses the cosmological horizon, flatness and monopole
problems and generates a nearly scale-invariant spectrum of density
perturbations without invoking superluminal expansion (inflation).
The scenario relies, instead, on physical phenomena that arise
naturally in theories based on extra dimensions and branes. As an example, we
present our scenario predominantly within the context of heterotic M-theory.
A  prediction that  distinguishes this scenario from standard
inflationary cosmology is a strongly blue gravitational wave spectrum,
which has consequences for  microwave background  polarization
experiments and gravitational wave detectors.

\end{abstract}

\section{Introduction} \label{intro}

The Big Bang model provides an accurate account of the
evolution of our universe from
 the time of nucleosynthesis until the
present, but does not 
address the key theoretical puzzles regarding the structure and make-up 
of the Universe, including:
{\it the flatness puzzle} (why is the observable universe so close to
being spatially flat?); 
{\it the homogeneity puzzle} (why are causally disconnected
regions of the universe so similar?);
{\it the inhomogeneity puzzle} (what is the origin of the density
perturbations responsible for the cosmic microwave background
anisotropy and large-scale structure formation? And
 why is their spectrum nearly scale-invariant?);
and {\it the monopole problem} (why are topological defects
from early phase transitions not observed?). 
Until now, the leading theory for resolving  these puzzles has been
the
inflationary model of the universe~\cite{Gut,Lin}. 
The central assumption of any 
inflationary model is that the universe
underwent a period of superluminal expansion early in its history before
settling into a radiation-dominated evolution. 
Inflation is
a remarkably successful theory. 
But in spite of twenty 
years of endeavor there is no convincing link 
with theories of quantum gravity such as
M-theory.

In this paper, we present a cosmological scenario 
which addresses the
above puzzles 
but which does not 
involve inflation.  
Instead, we invoke new physical phenomena that arise naturally 
in theories based on extra dimensions and branes. 
Known as ``brane universe'' scenarios, these ideas first appeared
in Refs.~\ref{valery} and~\ref{akama}. 
However, only recently were they given compelling motivation in the work
of Ho\v rava and Witten~\cite{witten1} and in the 
subsequent construction of heterotic
M-theory by Lukas, Ovrut and Waldram~\cite{lukas1}. 
Complementary motivation was
provided both in superstring theory~\cite{astrings,bstrings,cstrings} 
and in non-string contexts~\cite{RS,bns}. 
Many of the ideas discussed here are applicable, in principle,
to any brane universe theory. For example, in discussing the features of
our model, we draw examples
both from the Randall-Sundrum model~\cite{RS} and from heterotic
M-theory. However, in this paper, we emphasize heterotic M-theory. This is
done for specificity and because, by doing so, we know that we
are working in a theory that contains all the particles and interactions of the
standard model of particle physics. Hence, we are proposing a potentially
realistic theory of cosmology.

Specifically, our scenario  assumes a 
universe consisting of  a five-dimensional
space-time with two bounding (3+1)-dimensional surfaces (3-branes)
separated by a finite gap spanning an intervening bulk volume.
One of the boundary 3-branes (the ``visible brane'') corresponds to
the observed four-dimensional universe in which ordinary particles
and radiation propagate, and the other is a ``hidden brane."  
The universe 
begins as a cold, empty, nearly BPS
(Bogolmon'yi-Prasad-Sommerfield~\cite{BPS})
ground state of heterotic M-theory, 
as described by Lukas, Ovrut and Waldram~\cite{lukas1}. 
The BPS property is required in order to have a 
low-energy four-dimensional effective action with ${\cal N}=1$
supersymmetry.
The visible and hidden branes are flat (Minkowskian)
but 
the bulk is warped along the fifth dimension.

In addition to the visible and hidden branes, 
the bulk volume contains an additional 3-brane 
which is free to move across the bulk.  
The bulk brane may exist initially as a BPS state, or it may spontaneously
appear in the vicinity of the hidden brane through a process akin to
bubble 
nucleation. 
The BPS condition in the first case or the minimization of the action in
the 
second case require that the bulk brane be flat, oriented parallel to the 
boundary
branes, and initially at rest.
Non-perturbative effects result in a potential which attracts the bulk
brane 
towards the visible brane.
We shall assume that the bulk brane is much lighter 
than the bounding branes, so 
that its backreaction is a small correction to the geometry. 
See Figure 1.

The defining moment is the creation of 
the hot big bang universe by the collision of the slowly
moving bulk brane with our visible brane. Although the universe may exist
for an indefinite period prior to the collision, cosmic time as normally 
defined begins at impact.   The bulk and visible
branes fuse through a ``small instanton'' transition, during which a 
fraction of the kinetic energy of the bulk brane is converted
into a hot, thermal bath of 
radiation and matter on the visible brane. The universe
enters the hot big bang or Friedmann-Robertson-Walker (FRW)
phase.  Notably, instead
of starting from a cosmic singularity with infinite temperature, 
as in conventional big bang cosmology,
the hot, expanding universe in our scenario
starts its cosmic evolution at a finite temperature.
We refer to our proposal as the ``ekpyrotic universe'', a term drawn
from the Stoic model  of cosmic evolution
in which the universe
is  consumed by fire at regular intervals
and reconstituted out of this fire, a  
conflagration called ekpyrosis~\cite{volk}. Here, the universe 
as we know it is made (and, perhaps, has been remade) through 
a conflagration ignited by  collisions between branes along a hidden
fifth dimension.

Forming the hot big bang universe from colliding  branes
affects  each of the cosmological problems of the standard big
bang model.  First, the causal structure of space-time in the
scenario differs from the conventional big bang picture.
In standard big bang cosmology, two events separated by more than a
 few Hubble radii are causally disconnected. 
This relationship between the Hubble radius and the causal horizon 
applies to  FRW cosmologies that are expanding subluminally,          
as in the standard hot big bang model, but does not apply to 
more general cosmologies, such as de Sitter space-time or 
the scenario we will
describe. 
\begin{figure}
\begin{center}
{\par\centering \resizebox*{4in}{4in}{\includegraphics{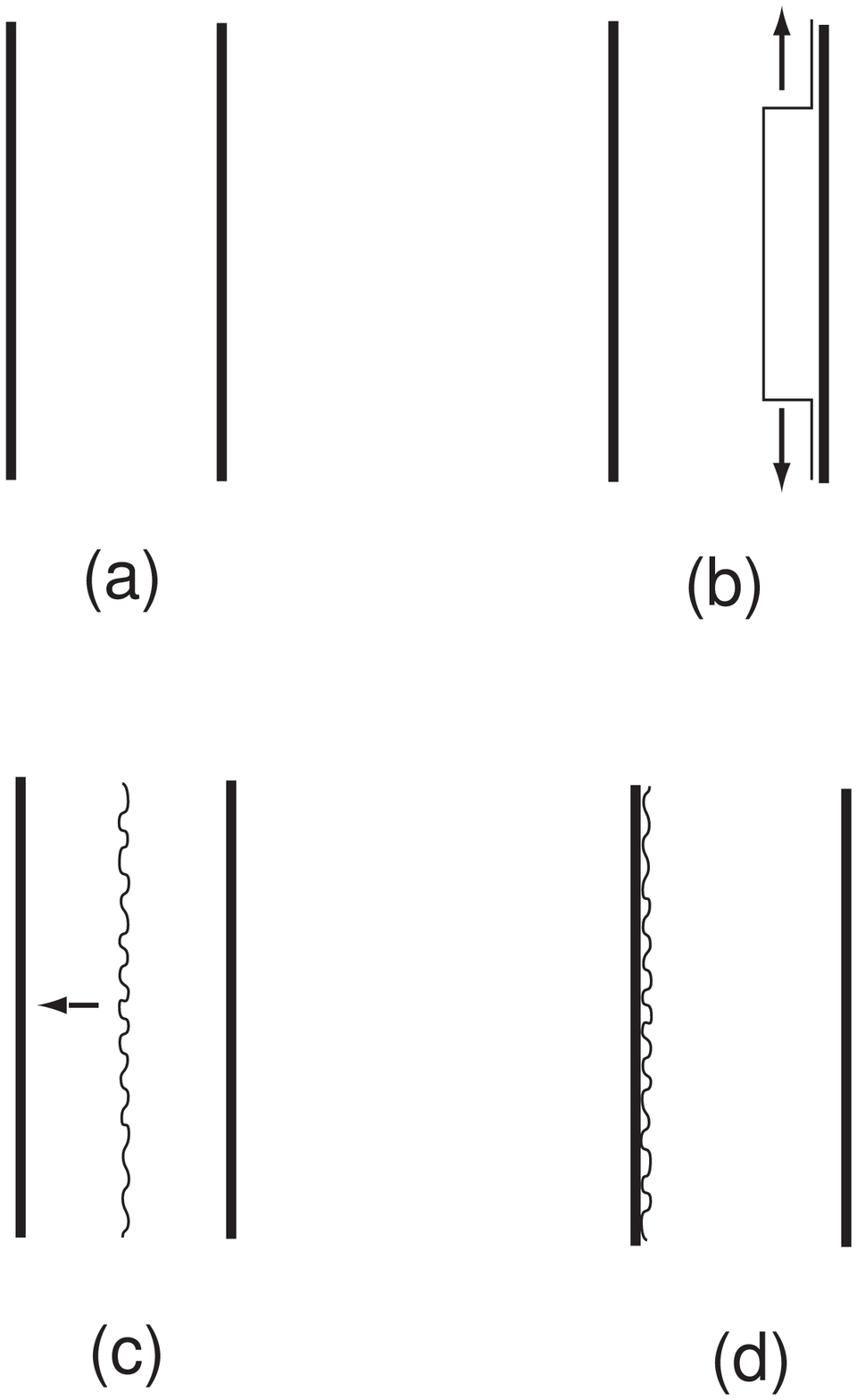}} \par}
\end{center}
\caption{One possible set of initial conditions of the ekpyrotic scenario has the universe 
beginning in a cold, vacuous,
nearly BPS state consisting of two static massive orbifold planes
and a warped geometry in the intervening bulk (a) in which
the curvature is low near the rightmost orbifold plane (the hidden
brane)
and high near the
leftmost orbifold plane (the visible brane).
Spontaneously, a bulk brane peels away from the hidden brane
over some region of space (b), forming a terrace. The edges of the terrace
expand outwards at light speed, while the interior moves very slowly
towards
the opposing visible brane. Although the bulk brane is flat on average,
quantum fluctuations produce ripples over a wide range of length scales
as the brane traverses the bulk (c). When the bulk brane collides with the
visible brane, the ripples result in different regions colliding and
reheating at slightly
different times (d),  thereby impressing a  spectrum of density
fluctuations
on the visible universe.
The energy from the collision is translated into matter and radiation,
heating the universe to a temperature a few orders of magnitude smaller
than the unification scale.}
\label{fig:terrace}
\end{figure}

\noindent
In our scenario, 
the collision sets the initial temperature
and, consequently, the Hubble radius at the beginning of the FRW 
phase. The Hubble radius is generally 
infinitesimal compared to the collision 
region. 
Two events outside the Hubble radius
are correlated since local conditions have
a common causal link,
 namely, the collision with the bulk brane.  
 Here we take advantage of the fact that the brane is a  
 macroscopic, non-local object
 and exists for an 
indefinitely long  period prior to the collision.
That is, there is no direct connection between the time transpired
preceding the collision or causality and the Hubble time.
This feature provides a natural means for resolving the
horizon problem.

As we have noted above, the boundary branes and the bulk brane are 
initially flat and parallel, as demanded by the BPS condition.
Furthermore, the motion of the bulk brane along the fifth dimension
maintains 
flatness (modulo small fluctuations around the flat background).
Hence, the hot big bang universe resulting from the collision of
a flat bulk brane with a flat visible brane 
is spatially flat. 
In other words, we address the flatness problem 
by beginning near a BPS ground state.

We do not require the initial state to be  precisely BPS to resolve
the horizon and flatness problems. It suffices if
the universe is  flat and homogeneous  on 
scales ranging up to the (causal) particle horizon, as should 
occur naturally beginning from more general initial conditions.
In the ekpyrotic scenario, the 
distance that particles can travel before 
collision is exponentially long because the bulk brane motion is 
extremely slow.  As a result, the particle horizon at collision can be
many more than 
60 e-folds larger than the Hubble radius at collision (where the latter 
is determined by the radiation temperature at collision).
That is, rather than introducing
superluminal expansion to resolve the horizon and 
flatness problems, the ekpyrotic model 
relies on 
the assumption that the universe began in
an empty, quasi-static BPS state which lasted
an exponentially long time 
prior to the beginning of the hot big bang phase.

Quantum fluctuations introduce ripples in the bulk brane as 
it moves across the fifth dimension. 
During this motion there is a scale above which
modes are frozen in, and below which they 
oscillate. This scale decreases with time in a manner akin to the
Hubble radius of a collapsing Universe.
The fluctuations span all scales up to this freeze-out scale,
and we assume they begin in their quantum mechanical ground state.
As the brane moves across the bulk, the effective Hubble radius
shrinks
by an exponential factor while the wavelengths of the modes decrease only 
logarithmically in time. 
Consequently, modes that begin inside the initial freeze-out scale
end up exponentially far outside it, and 
exponentially far outside the final Hubble radius 
at the time of collision.  
The ripples in the bulk brane cause the collision between the 
bulk brane and the visible brane to occur at slightly
different times in different regions of space.
 The time differences mean regions heat 
 up and begin to cool at different times, 
resulting in adiabatic temperature and density fluctuations.  
Hence, as interpreted by an observer in the hot big bang FRW phase,
the universe begins with a spectrum of density fluctuations that extend to
exponentially large, super-horizon scales.

Of course, 
the spectrum of energy density perturbations 
must be nearly scale-invariant (Harrison-Zel'dovich~\cite{HZ})
to match observations of large-scale structure formation
and temperature anisotropies in the cosmic microwave background.
We find that this condition is 
satisfied if the potential 
attracting the bulk brane towards the visible brane is ultra
weak at large separations. 
This is consistent with potentials generated by
non-perturbative effects such as
the exchange of virtual M2-branes 
(wrapped on holomorphic curves) between the bulk brane and either of the
boundary branes.

While the spectrum of perturbations is approximately scale invariant,
as with inflation there are small deviations from scale invariance.
In the examples considered here,
the spectrum is 
blue (the amplitude increases as the wavelength decreases),
in contrast to typical inflationary models.  With  exponentially
flat potentials,
the spectrum is only marginally blue, consistent with current 
observations.  On the other hand, the potential has no effect
on the tensor (gravitational wave) perturbations, so the tensor 
spectrum
is strongly blue (spectral index $n_T \approx 2$), in contrast to
the  slightly red ($n_T \le 0$) spectrum predicted in most
inflationary models. 
This prediction may be tested in
near-future microwave background anisotropy and gravitational wave
detector
experiments.

For some aspects of the ekpyrotic scenario, such as the generation of
quantum 
fluctuations, the description from the point-of-view of an observer 
on the bulk brane is the most intuitive.
For that observer, the scale factor and Hubble radius appear to be 
shrinking because the warp factor decreases as the bulk brane moves across
the fifth dimension.
However, as observed from the near-stationary boundary orbifold planes,
the universe is slowly expanding
due to the gravitational backreaction caused by the bulk brane
motion. Indeed, a feature of the scenario is that
the bulk brane is responsible for initiating
the expansion of the boundary branes. 
Furthermore, as we  shall show,
the brane gains
kinetic energy due to its coupling to  moduli fields.
Upon impact, the bulk brane is absorbed by the visible 
brane in a so-called small-instanton phase transition (see Sec. II for
details). 
This transition can change the gauge group on the visible brane. (For example,
before collision the gauge group on the visible 
brane might be one of high symmetry, such as $E_6$, whereas after collision it becomes the standard
model gauge group.)
Furthermore, the number of light families of quarks and leptons on the visible brane may 
change during the transition. (For example, the visible brane might make a transition from 
having no light families of quarks and leptons to having three.) 
Upon collision,
the kinetic energy gained by the bulk brane  is converted
to thermal excitations of the light degrees of freedom, and the hot
big bang phase begins. Hence, the brane collision is not only responsible
for
initiating the expansion of the universe, but also for spontaneously
breaking symmetries and for producing all 
of the quarks and leptons.

If the maximal temperature lies well below the mass scale of 
magnetic monopoles (and any other cosmologically dangerous 
 massive, stable particles or defects),
none will be generated during the collision and the monopole problem 
is avoided.

Although the ideas presented in this work may be applicable to more
general 
brane-world scenarios,
such as Randall-Sundrum, 
in developing our scenario
we have felt that  it is important to take 
as a guiding principle that
any concepts introduced in this scenario be consistent with 
string theory and M-theory. 
By founding the model on concepts from heterotic M-theory, 
one knows from the outset that the theory is 
rich enough to contain the particles and symmetries necessary 
to explain the real universe and that nothing we introduce 
interferes with a fundamental theory of quantum gravity.
We emphasize that our scenario does not rely on exponential
warp factors (which are inconsistent with BPS ground states in 
heterotic M-theory) 
nor does it require large  (millimeter-size) extra dimensions.
For example,
we consider here  a bulk space whose size is only four or five
orders of magnitude
larger than the Planck length, consistent with Ho\v rava-Witten
phenomenology~\cite{witten2,dine}.
All brane universe
theories, including heterotic M-theory, suffer from some poorly understood
aspects. For example, we have nothing to add here about the stability of
the final, late time vacuum brane configuration. We will simply assume
that
branes in the early universe move under their respective forces until,
after
the big bang, some yet unknown physics stabilizes the vacuum.

Figure 1 summarizes the conceptual picture for one possible set of initial conditions.
The remainder of this
paper discusses our attempt 
to transform the conceptual framework into a concrete
model.  For this purpose, a number of technical advances 
have been  required:
\begin{itemize}
\item an understanding of the perturbative BPS ground state (Section II)
and  how it can  lead to the initial conditions
desired for our scenario (Section III);
\item a moduli space formulation of brane cosmology (Section IV A);
\item a derivation of the equations of motion describing the propagation
of bulk branes in a warped background in heterotic M-theory, including
non-perturbative effects 
(Section IV B-D);
\item a computation of the bulk brane-visible brane collision energy,
which sets the initial temperature  and expansion rate
of the FRW phase (Section IV D);
\item an analysis  of how  the gravitational backreaction due to
the motion of the bulk brane   induces
the initial expansion of the universe (Section IV E);
\item  a computation of how ripples in the bulk brane  translate into
 density perturbations after collision with the
  visible brane
(Section V A);
\item a theory of how quantum fluctuations produce
a spectrum of ripples on the bulk brane as it propagates
through a warped background (Section V B);
\item a determination of the generic conditions for obtaining a nearly
scale invariant spectrum and
application of general principles  to designing specific
 models
 (Section V B);
 \item a calculation of the tensor (gravitational wave) perturbation
 spectrum (Section V C);
\item a recapitulation of the full scenario explaining how 
the different components rely on properties of moduli in 5d (Section VI A);
\item a fully worked example which satisfies all cosmological constraints
(Section VI A);
\item and, a comparison of the ekpyrotic scenario
with inflationary cosmology, especially differences in their 
predictions for the fluctuation spectrum (Section VI B).
\end{itemize}
In a subsequent paper, we shall elaborate on the moduli space 
formulation of brane cosmology and show how it leads to a novel
resolution of the singularity problem of big bang cosmology~\cite{future}.

The ekpyrotic proposal bears some relation to
the pre-big bang scenario of 
Veneziano {\it et al.}\cite{Venez,Gasp,APT} 
which begins with an almost empty but unstable vacuum state of 
string theory but which, then, undergoes superluminal deflation.
Several important conceptual differences are discussed in Section~VIB.
Models with brane interactions that drive inflation followed by brane collision have
also been considered.\cite{dvali1,dvali2,dvali3,Park,Alex}
Applications of the moduli space of M-theory and Ho\v rava-Witten theory
to cosmology have been explored previously in the context of inflation~\cite{banks,lukas4,lukas5}.
The distinguishing feature of the ekpyrotic model is that it avoids inflation
or deflation altogether.
A non-inflationary solution to the horizon problem was suggested 
in  Ref.~\ref{stark}, but it is not clear how to generate a 
nearly scale-invariant spectrum of density fluctuations
without invoking inflation.

We hope that our technical advances may be useful in exploring
other variants of this scenario.  Some aspects remain more speculative,
especially the theory of initial conditions (as one might expect)
and non-perturbative contributions.  A detailed understanding of 
these latter aspects awaits progress in heterotic M-theory.

\section{Terminology and Motivation from Heterotic M-theory}

In this section, we briefly recount key features of heterotic M-theory
that underlie and 
motivate the example of our ekpyrotic scenario given in this
paper. Those who wish to understand the basic cosmological scenario 
without regard to the heterotic
M-theory context can proceed directly to the next section.

Heterotic M-theory has its roots in the work of 
Ho\v rava and Witten~\cite{witten1} who
showed that compactifying (eleven-dimensional)
M-theory on an $S_1/Z_2$ orbifold corresponds to the strong-coupling limit of 
heterotic $E_8\times E_8$ (ten-dimensional) string theory.
Compactifying an additional six dimensions on a Calabi-Yau three-fold leads, 
in the low energy limit,
to a four-dimensional ${\cal N}=1$ supersymmetric theory~\cite{witten2},
the  effective field theory that underlies many supersymmetric theories of
particle phenomenology.
By equating the effective gravitational and grand-unified coupling
constants to their physical values, 
it was realized that the orbifold dimension is 
a few orders of magnitude
larger than the characteristic size of the Calabi-Yau 
space~\cite{witten2,dine}.
There is, therefore, a substantial energy range over which the universe is
effectively five-dimensional, being bounded in the fifth-dimension by two
$(3+1)$-dimensional 
``end-of-the-world'' $S^{1}/Z_{2}$ 
orbifold fixed planes. By compactifying Ho\v rava-Witten theory on a
Calabi-Yau three-fold, in the presence of a non-vanishing $G$-flux 
background (that is, a non-zero field strength of
the three-form of eleven dimensional M-theory)
required by anomaly cancellation, the authors of Refs.~\ref{lukas1}
and~\ref{lukas2} 
were able to derive the
explicit effective action describing this five-dimensional regime.
This action is a specific gauged version of ${\cal N}=1$ supergravity in five
dimensions 
and includes ``cosmological'' potential terms that always arise in
the gauged context. 
It was shown in Ref.~\ref{lukas1} that these potentials support BPS
3-brane solutions of the equations of motion, the minimal vacuum
consisting of two 3-branes, each coinciding
with one of the $S_1/Z_2$ orbifold fixed planes. 
These boundary 3-branes (the visible brane and the hidden brane) each inherit a
(spontaneously broken) ${\cal N}=1$ $E_8$ super gauge multiplet from Ho\v rava-Witten
theory. This five-dimensional effective theory with BPS
3-brane vacua is called {\it heterotic M-theory}. 
It is a fundamental paradigm for ``brane universe'' scenarios of particle 
physics.

In order to support a realistic theory, heterotic M-theory must include
sufficient gauge symmetry and particle content on the 3-brane boundaries.
In the compactification discussed above, the authors of Refs.~\ref{lukas1}
and~\ref{lukas2} initially made use of 
the standard embedding of the spin connection into 
the gauge
connection, leading to an $E_6$ gauge theory on the visible 3-brane while the 
gauge
theory on the hidden 3-brane is $E_8$.
Subsequently, more general embeddings than the standard one were
considered~\cite{lukas3}. 
Generically, such non-standard embeddings 
(topologically non-trivial configurations of gauge fields known as
$G$-instantons) 
induce
different gauge groups on the orbifold fixed planes. 
For example, it was shown
in Refs.~\ref{donagi} and~\ref{pantev} that one can obtain grand unified gauge 
groups such as $SO(10)$, $SU(5)$
and the standard model gauge group $SU(3)_{C} \times SU(2)_{L}\times U(1)_{Y}$
on the visible brane by appropriate choice of $G$-instanton.
Similarly, one can obtain smaller gauge groups on the hidden brane, such as $E_7$ and $E_6$.
However, as demonstrated in Refs.~\ref{lukas3}-\ref{pantev}, 
requiring a physically interesting gauge group
such as $SU(3)_{C} \times SU(2)_{L} \times U(1)_{Y}$, along with the requirement 
that
there be three families of quarks and leptons, typically leads to the constraint that
there must be a certain number of M5-branes in the bulk space
in order to make the theory anomaly-free.
These M5-branes are wrapped on holomorphic curves in the Calabi-Yau
manifold, and appear as 3-branes in the five-dimensional effective
theory. 
The five-dimensional effective action for non-standard embeddings and bulk
space M5-branes was derived in Ref.~\ref{lukas3}.
The key conclusion from this body of work is that heterotic M-theory
can incorporate the particle content and symmetries required for a 
realistic low-energy effective theory of particle phenomenology.
There has been a considerable amount of literature
studying both the four-dimensional limit of Ho\v rava-Witten
theory~\cite{hw} and heterotic M-theory~\cite{het}.

A  feature of M5-branes which we will utilize 
is that they are allowed to move along
the orbifold direction. 
It is important to note that the
M5-brane motion through the bulk is not ``free'' motion. Rather,
non-perturbative effects, such as the exchange of virtual 
open supermembranes stretched between a 
boundary brane and the bulk M5-brane, can produce a force between them. 
The corresponding potential energy can, in principle, be computed
from M-theory. Explicit calculations of the supermembrane-induced
superpotentials in the effective four-dimensional theory have been carried out
in Refs.~\ref{moore} and~\ref{lima}. 
Combined with the M5-brane contribution to the K\"{a}hler potential
presented in Ref.~\ref{derrin}, one can obtain an expression for the supermembrane 
contribution to the M5-brane potential energy.
These non-perturbative effects cause the bulk brane to move along the extra 
dimension.
In particular, it may come into contact with one of the orbifold fixed planes. 
In this case, the branes undergo a ``small instanton'' phase transition
which effectively dissolves the M5-brane, absorbing its ``data'' into the
$G$-instanton~\cite{park}. 
Such a phase transition at the visible brane may change the number of 
families of quarks and leptons as well as the gauge group. 
For instance, the observable 3-brane may go from having no 
light families of quarks and leptons
prior to collision to having three families after the phase transition.

It is in this five-dimensional world of heterotic M-theory, bounded at the
ends of the fifth dimension by our visible world and a hidden world, and
supporting moving five-branes subject to catastrophic family and gauge
changing collisions with our visible 3-brane, that we propose to find a
new theory of the very early universe.

\section{Initial conditions} \label{init}

All cosmological models, including the hot big bang and the 
inflationary scenario, 
rest on assumptions about the initial conditions. Despite
attempts, no rigorous theory of initial conditions yet exists. 
%

Inflationary theory conventionally 
 assumes that the universe emerges in a high energy state of
no particular symmetry that is rapidly expanding.
If traced backward in time,  such states possess an
initial singularity. This is only one of 
several fundamental obstacles to 
constructing a well-defined theory of `generic' inflationary 
initial conditions. For a theory based on 
such general and uncertain initial conditions, it is essential 
that there be a dynamical attractor mechanism that makes the universe
more homogeneous as expansion proceeds, since such a mechanism 
provides hope that the uncertainty associated with
the initial conditions is, in the end, irrelevant.  Superluminal expansion 
provides that mechanism.

The ekpyrotic model is 
instead
built on the assumption that the initial state
is quasi-static, nearly vacuous and long-lived, 
with properties dictated by symmetry. 
So, by construction, the initial 
state is special, both physically and mathematically.
In this case, while a dynamical attractor mechanism
may be possible (see below), it is not essential.  One can 
envisage the possibility that the initial conditions are simply the
result of a selection rule dictating maximal
symmetry and nearly zero energy. 

Within the context of superstring theory and M-theory, a natural
choice with the above properties
is the BPS (Bogolmon'yi-Prasad-Sommerfeld) state.\cite{lukas1}
The BPS property is already
required from  particle physics in order to have a
low-energy, four-dimensional
effective action with ${\cal N}=1$ supersymmetry,
necessary for  a realistic
 phenomenology.  For our purposes,
  the BPS state is ideal because,
  not only is it homogeneous,  as one might suppose,
  but it is also flat.  That is, the BPS condition
  links curvature and  homogeneity. 
  It requires the two boundary branes to be parallel.

The BPS condition also requires the bulk brane to be nearly stationary.  
If the bulk brane has a small initial velocity, it is free
to move along the fifth dimension. Assuming the bulk brane to 
be much lighter than the bounding branes enables us to treat 
its backreaction  as a small perturbation
on the geometry. 
Non-perturbative effects can modify this picture by introducing 
a potential for the bulk brane.  For example, a bounding 
brane and bulk brane can interact by the exchange of M2-branes
wrapped on holomorphic curves, resulting in a potential drawing
the bulk brane towards our visible brane.
In Section V, we shall see that the non-perturbative potential plays 
an important role in determining the spectrum of energy density
fluctuations following the collision of the bulk brane with our
visible brane.

We do not rule out the possibility of a dynamical attractor mechanism
that drives the universe towards the BPS state beginning from some 
more general initial condition.  Such parallelism would be a natural
consequence of all the branes emerging from one parent brane.
Another appealing  possibility is to begin with a  configuration
consisting of only the two bounding 3-branes.  
The  configuration may have some curvature and ripples,  
but these can be dissipated by radiating excitations tangential
to the branes and having them travel off to infinity.  
 Then, at some instant, the hidden brane 
may under go a ``small instanton" transition which causes
a bulk brane to peel off.  While little is  known 
about the dynamics of this peeling off, it is reasonable to 
imagine a process similar to bubble nucleation in first
order phase transitions.  
We suppose that there is a long-range, attractive, non-perturbative
potential that draws the bulk brane towards our visible brane,
as shown in Figure 2.
Very close to the hidden brane, there may be a short-range
attractive force between the bulk brane and the hidden brane due to 
small-instanton physics (not shown in the Figure). 
The situation is similar to false vacuum decay where the position of the 
bulk brane along the fifth dimension, $Y$, plays
the role of the order parameter or scalar field.
Classically, a bulk brane attached to the hidden brane is 
kept there by the energy barrier. However, quantum mechanically,
it is possible to nucleate a patch of brane for which $Y$ lies on the 
other side of the energy barrier. At the edges of the patch, the
brane stretches back towards and joins onto the orbifold plane.
The nucleated patch would correspond to the minimum action
tunneling configuration. We conjecture that, as in the case of 
bubble nucleation, the configuration would be one with maximal
symmetry. In this case, this configuration would correspond to a patch 
at fixed $Y$ with a spherical boundary along the transverse 
dimensions. The nucleation would appear as the spontaneous
appearance of a brane that forms a flat, spherical terrace
at fixed $Y$ parallel to the bounding branes ({\it i.e.}, flat).
Once nucleated, the boundaries of the terrace would spread
outwards at the speed of light, analogous to the outward 
expansion of a bubble wall in false vacuum decay.
 At the same time,  the brane (the terrace) would
travel towards the visible brane due to the non-perturbative
potential.
As we shall see in Section~\ref{motion}, this motion is very slow
(logarithmic with time) so that the nucleation rate is essentially
instantaneous compared with the time scale of transverse motion.
In other words, for our purposes, the nucleation process corresponds to a
nearly infinite brane peeling off almost instantaneously.

Beginning in an empty,  quasi-static  state addresses the
horizon and flatness problems, but
one should not underestimate the
remaining challenges:
how to generate a hot universe, and how to generate
perturbations required for large-scale
structure. 
The remarkable feature of  the ekpyrotic picture, as shown in 
the forthcoming sections, is that brane collision naturally
serves both roles.

\begin{figure}
{\par\centering \resizebox*{4in}{4in}{\includegraphics{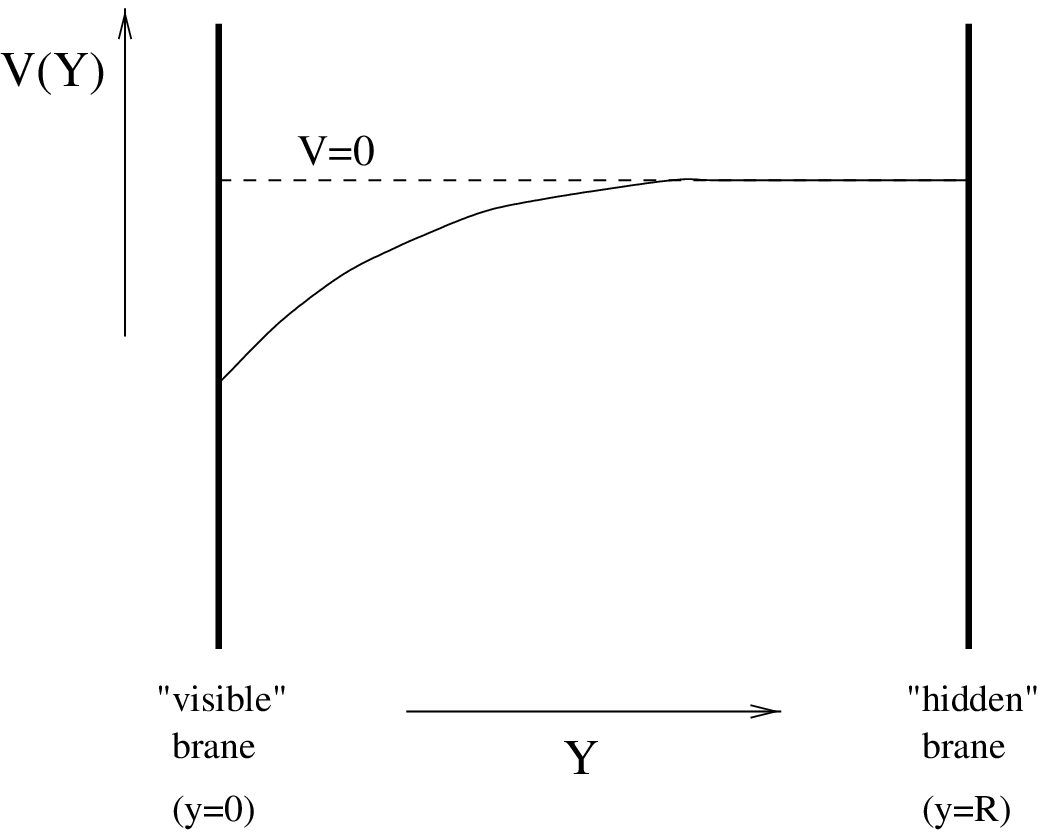}} \par}
\caption{Sketch of the exponential potential $V(Y)=-ve^{-m \alpha Y}$ 
(the line of zero
potential energy corresponds to the dotted line). The potential attracts the
bulk brane towards the visible brane. 
The force is strongest near
the visible brane and tends to zero at large distances.}
\label{fig:V}
\end{figure}

In either setup, we begin with a flat bulk brane, either 
a finite patch or an infinite plane, which starts nearly at 
rest, and a non-perturbative
force drawing it towards the visible brane. These initial conditions are
sufficient to enable our scenario.

\section{Propagation of the Bulk Brane} \label{motion}

\subsection{Moduli Space Actions for Brane-World Gravity} \label{moduli}

In this section, we discuss the moduli space approximation which
we shall employ throughout our analysis.  
This approximation may be used when there is 
a continuous family of static solutions of the field 
equations, of degenerate action. It is the basis for much of what is
known of the classical and quantum  properties of
solitons such as magnetic monopoles and
vortices \cite{monopoles}. It is also
a powerful tool for cosmology as it neatly
approximates the 
five dimensional theory in the regime where
the rate of change of the geometry, as measured, for example, by the 
four dimensional 
Hubble constant $H$ is smaller than the
typical spatial curvature scale in the static solutions. 
The moduli are the 
parameters specifying the family of static solutions,
`flat directions' in configuration space
along which slow dynamical evolution is
possible. During such evolution the excitation
in other directions is consistently small
provided those
directions are stable 
and characterized by large oscillatory 
frequencies.

The action on moduli space is obtained by
substituting the static solutions into the full action
with the modular parameters represented as
space-time dependent
moduli fields, $Q^I$, where $I$ runs over all the moduli fields. 
If we consider the time dependence 
first, as we shall do for the homogeneous background solutions,
the moduli space action takes the form
\begin{eqnarray}
{\cal S} = \int d\t \, G_{IJ}(Q) \dot{Q^I} \dot{Q^J}
\label{eq:m0}
\end{eqnarray}
where $G_{IJ}(Q)$ is a matrix-valued function of the 
moduli fields.
This is the action for
a non-relativistic particle moving in a background metric
$G_{IJ}(Q)$, the metric on moduli space.
For truly degenerate static solutions, the potential term
must be constant, and therefore irrelevant to the dynamics.
Even if a weak potential $V$ is additionally present, as it shall be in our
discussion below,
the moduli approximation is still valid
as long as 
the dynamical evolution consists 
in the first
approximation of an adiabatic progression through the space of
static solutions. 

In the next section we compute the moduli space action
for heterotic M-theory. First, however, it is instructive to consider a 
simpler model which demonstrates similar
physical effects and is of some
interest in its own right. This model, discussed by 
Randall and Sundrum \cite{RS}, consists of a five dimensional bulk
described by Einstein 
gravity with a negative cosmological constant $\Lambda$, bounded 
by a pair of branes with 
tension $\pm \alpha$. The brane tension must be
fine tuned to the value 
$\alpha=(3|\Lambda|/(4\pi G_5))^{1/2}$ in order for 
static solutions to exist. Because 
there are few moduli,
the model is simple to analyze and illustrative of some
important effects. However, as we emphasize below,
there are reasons for taking heterotic M-theory 
more seriously as a candidate fundamental description.

In the Randall-Sundrum model, the static field equations allow
a two-parameter family of solutions, with metric
\begin{eqnarray}
ds^2= -n^2 d\t^2 +a^2 d\vec{x}^2+ dy^2,\quad 
a=e^{{y/ L}}, \quad n=N a, \qquad y_1\leq y \leq y_0,
\label{eq:m1}
\end{eqnarray}
and the positive (negative) 
tension branes located at $y_0$ ($y_1$) respectively. 
$L$ is the Anti-de Sitter (AdS) radius 
given by $L^2= 3/( 4\pi G_5 |\Lambda|)$,
and 
$N$ is an arbitrary constant.

The above solutions are specified by the three moduli $N$, 
$y_0$ and $y_1$. We now allow them to be 
time dependent. The lapse function $N(t)$ is associated
with time reparametrization invariance and, hence,
not a physical degree
of freedom. The other two moduli represent the proper distance
between the branes, but also the time-dependent cosmological 
`scale factors' 
$a_0
= e^{y_0 /L}$ and $a_1 = e^{y_1 /L}$ on each brane. 
We shall see in a moment how
these combine to give four-dimensional gravity with a massless
scalar, related to the proper separation between the branes.

To compute the moduli space action it is 
convenient to change from the coordinates in
Eq.~(\ref{eq:m1}) to coordinates in which the branes are fixed.
This is accomplished by setting 
$y=(y_0-y_1)\tilde{y}
+y_1$, and the branes are now located at $\tilde{y}=0$ and
1 respectively.
We substitute the ansatz Eq.~(\ref{eq:m1}) into the 
five dimensional action and integrate over $\tilde{y}$.
All potential terms cancel. Since the branes do not move in 
the $\tilde{y}$ coordinates, the kinetic terms arise only from
the five dimensional Ricci scalar, which yields the 
result
\begin{eqnarray}
{\cal S} 
={ L\over 16 \pi G_5} \int d\t d^3x N^{-1} \, 6 (\dot{a}_1^2 -\dot{a}_0^2).
\label{eq:m13}
\end{eqnarray}
Thus the  metric on moduli space is just the 1+1 Minkowski
metric, and we infer that moduli space in this theory is
completely flat.

 We may now change coordinates to
\begin{eqnarray}
a_0=a \,\cosh f \qquad a_1=a \,\sinh f
\label{eq:m13a}
\end{eqnarray}
where  $a^2=a_0^2-a_1^2$, and the proper separation between the branes
is just $L\ln(a_0/a_1)=L\ln(\coth f)$. The action (\ref{eq:m13}) then becomes 
\begin{eqnarray}
{\cal S}
={ 1\over 16 \pi G_4} \int d\t d^3x  N^{-1} \, 6 
\left(-\dot{a}^2 +a^2\dot{f}^2 \right),
\label{eq:m13b}
\end{eqnarray}
where $G_4=G_5 L^{-1}$,
just the action for four dimensional gravity coupled to a massless scalar
field $f$. This is the `radion' field
 \cite{binrad}. Note that the `4d Einstein frame scale
factor' $a$ is {\it not} the scale factor that is seen by matter
localized on the branes: such matter sees $a_0$ or $a_1$. 
As we shall see, it is perfectly possible for both of the latter 
scale factors to expand while $a$ contracts.

The general solution to the moduli space theory is easily obtained
from (\ref{eq:m13}): the scale factors evolve linearly
in conformal time $\t$, with $\dot{a}_1=\pm \dot{a}_0$.
From the point of view of each brane, the motion of the other
brane acts as a density of radiation (i.e., allowing $\dot{a}_{1,0}=$ 
constant). Although the 
moduli space theory has identical local equations to four-dimensional
gravity coupled to a massless scalar, the geometrical interpretation
is very different, so that what is singular from one point of
view may be non-singular from the other~\cite{future}.

To obtain an action that might describe the present, hot big bang phase with
fixed gravitational constant,
one might add a
potential
$V(f)$ which fixes the interbrane separation
$f$, causing it to no longer be a free modulus (e.g., see Ref.~\ref{gold}). 
Likewise adding extra fields on either brane, 
one sees that a standard Friedmann constraint is obtained
from the variation with respect to $N$. Generally the matter
couplings will involve the field $f$. However, if $V(f)$ rises
steeply away from its minimum and if its value at the minimum
is zero (so that there is no vacuum energy contribution),
$f$ will not evolve appreciably and four dimensional
gravity will be accurately reproduced. 

We imagine that, prior to the big bang phase,
 there is an additional,
positive tension bulk brane.
For static solutions to occur, we require that
the three brane tensions 
sum to zero. 
Positivity of the bulk brane tension imposes that 
the cosmological constant 
to its right, $\Lambda_0$, be smaller in magnitude 
than that to its left, $\Lambda_1$, so that the 
corresponding 
AdS radii  obey $L_0>L_1$.

For the three brane case we obtain the moduli space action
\begin{eqnarray}
{\cal S} = 
{1\over 16\pi G_5} \int d\t d^3x N^{-1} 
6\left(-L_0\dot{a}_0^2 +
(L_0-L_1)\dot{a}_B^2 +L_1\dot{a}_1^2
\right),
\label{eq:m14}
\end{eqnarray}
where $a_B$ is the scale factor on the bulk brane.
Again this is remarkably simple, just 
Minkowski space of
2+1 dimensions. Likewise 
for $N$ parallel branes, all with positive tension except the negative
tension boundary brane, 
the system possesses a metric which is that for 
$N$-dimensional Minkowski
space. Note that the `masses' appearing in the kinetic
terms for the boundary branes are the opposite of what 
one might naively expect. Namely, the positive tension
boundary brane (hidden brane) has the negative `mass' $-L_0$, whereas
the negative tension boundary brane (visible brane) has a positive `mass' $L_1$.
The magnitudes of these terms are also surprising:
the visible brane has the greater magnitude tension, but the smaller magnitude 
moduli space mass.

Just as for the brane-antibrane system, 
when more branes are present one can 
change variables to those in which the theory resembles
four dimensional Einstein gravity coupled
to massless fields. For three branes, the required change of
variables is 
\begin{eqnarray}
a_0&=&a \,{\rm cosh} f, \cr
a_B&=&(L_0/(L_0-L_1))^{{1\over 2}} 
\,a \,{\rm sinh} f \,{\rm cos} \theta , \cr
 a_1&=&(L_0/L_1)^{{1\over 2}} 
\,a \,{\rm sinh} f \,{\rm sin} \theta, 
\label{eq:m14c}
\end{eqnarray}
and 
the scalar field kinetic term takes the form
$a^2 (\dot{f}^2  + \dot{\theta}^2\sinh^2 f )$. The 
scalar fields
live on the hyperbolic plane $H^2$, and there is a nontrivial 
K\"ahler potential.
When we introduce potentials, we must do so in a manner which respects
four dimensional general coordinate invariance. This restricts the 
form to 
\begin{eqnarray}
\Delta {\cal S} = -
\int d \t d^3 x N a^4 V(f,\theta).
\label{eq:m14p}
\end{eqnarray}

\begin{figure}
 \begin{center}
{\par\centering \resizebox*{4in}{4in}{\includegraphics{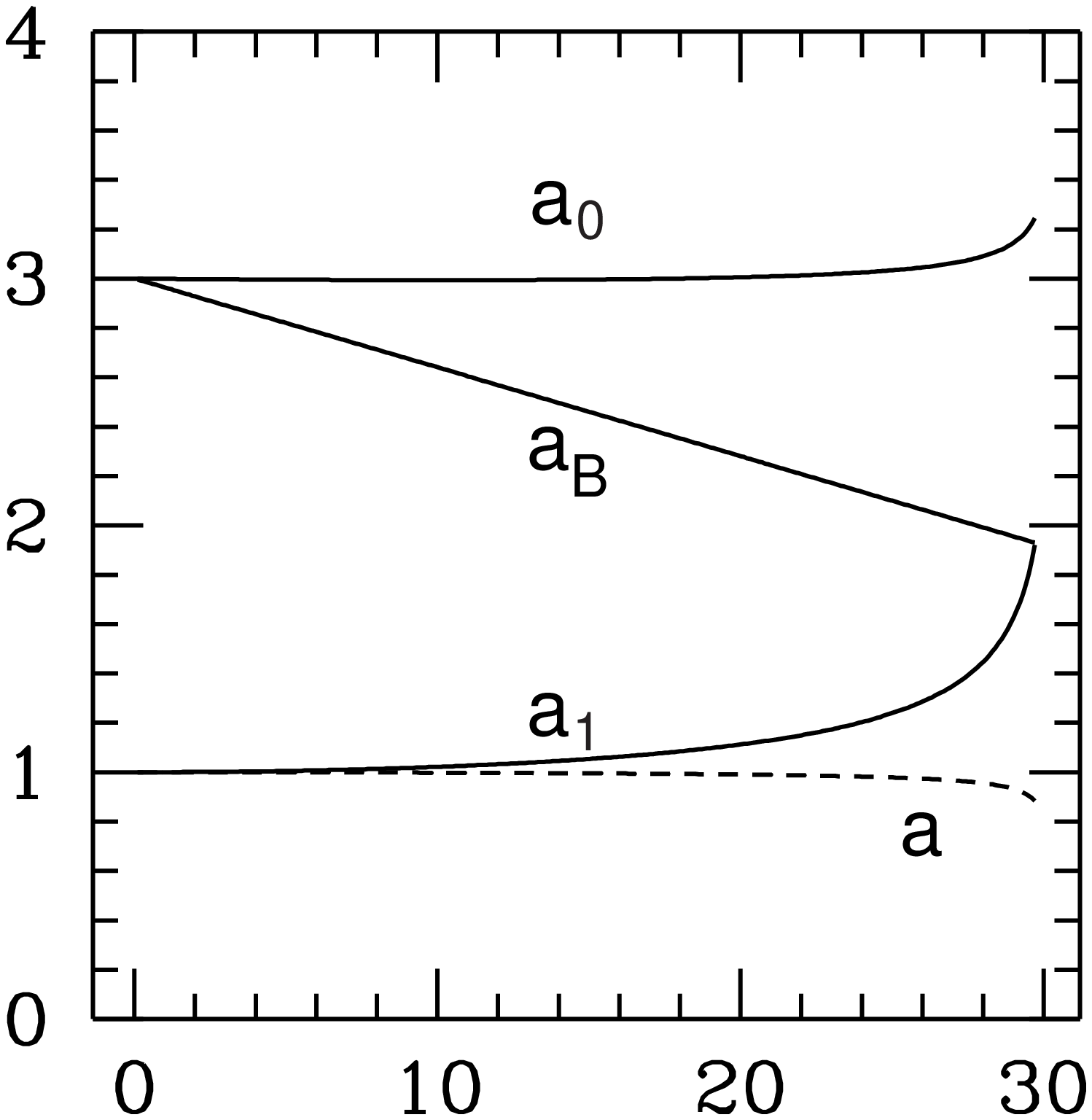}} \par}
  \end{center}
  \caption{Evolution of scale factors in a three-brane system where
  a bulk brane is drawn across from the hidden brane to the visible brane. The solid lines
show (from top to bottom) the scale factors on the hidden brane ($a_0$), the bulk brane ($a_B$) 
and the visible brane ($a_1$).
At collision between the bulk brane and the visible brane, both boundary branes are expanding. The evolution of
the 4d Einstein frame scale factor $a$ is also shown
as a dashed line:
it contracts throughout.
  }
  \label{fig:collads}
  \end{figure}

We shall assume that the system starts out nearly static,
and that the potential energy is always negative,
so that we are clearly not using inflation to drive expansion.
We further assume that the interaction
potential draws the bulk brane away from the hidden brane
towards the visible brane.
The original variables in Eq.~(\ref{eq:m14})
provide some insight into what happens. We consider an
interaction between $a_0$ and $a_B$ causing the latter to decrease. 
But since $a_0$ has a negative `mass'
it is actually pushed in the same direction and thus contracts.
Similarly an interaction potential between $a_0$ and $a_1$ can have the
opposite effect: causing both $a_0$ and $a_1$ to expand. 
Figure (\ref{fig:collads}) shows an example, 
with the bulk brane being pushed across the gap, and
the visible brane being attracted towards it.
The hidden brane actually `bounces' (this is barely visible in the Figure) 
due to a competition
between the two effects, so that by the collision
between bulk and visible branes, both outer boundary
branes are actually expanding; that is, $a_0$ and $a_1$ are both
increasing.   At first sight this
appears  inconsistent with the four dimensional point of view:
if the system starts out static, and with
all potentials negative, then the 4d Einstein frame
scale factor $a$ must contract throughout. 
The two points of view are consistent because 
$\dot{a}_1$ contributes negatively to
$\dot{a}$. Since $a_1$ is expanding rapidly compared to $a_0$,  $a$
is indeed contracting, as shown in the Figure. Of course what is
happening physically is that the fifth dimension is collapsing, 
a well known hazard of
Kaluza-Klein 
cosmology.
Here, the inter-brane separation is decreasing 
while the scale factors seen by matter on the branes
are expanding.

What happens when the bulk brane meets the boundary?
A matching condition is needed to determine the resulting 
cosmology.  
The initial
state is specified by the scale factors 
$a_0$, $a_B$ and $a_1$ and their
time derivatives, the final state by $a_0$, $a_1$
and their time derivatives, plus the coordinates
and momenta of any excitations produced in the collision.

One expects that the scale factors 
$a$, $a_1$ and
$a_0$ should be  continuous: likewise $\dot{a}_0$ would
be expected to be continuous if no bulk brane hits
the hidden brane, and if no constraint on the 
size of the extra dimension is imposed. 
However, $\dot{a}_1$ cannot be continuous since the bulk brane
imparts some momentum on the visible brane.
The momentum conservation condition can be expressed as 
\begin{eqnarray}
 L_0 a^i_0 \dot{a}_0^i -(L_0-L_1) \dot{a}_B^i a_B^i 
-L_1 \dot{a}_1^i a_1^i
=  L_0 a_0^f \dot{a}_0^f-L_0a_1^f\dot{a}_1^f,
\label{eq:m14pc}
\end{eqnarray}
where $i$ and $f$ subscripts label initial and final quantities.
Conservation of momentum applies if
the forces derive from a potential $V$ which is
short-ranged and translation invariant  and if there is no 
other entity that carries momentum after collision.
The above conditions imply the continuity of the
`4d Einstein frame scale factor' $a$ and its time derivative.

This matching condition, if correct,
would pose a serious problem for our scenario
since it implies that the four dimensional scale factor $a$
is contracting after collision, at the beginning of the hot 
big bang phase.
Specifically, the matching condition
suggests a simple continuation of the 
motion in Figure~\ref{fig:collads} after collision in which 
both branes are expanding  but $a$ is 
decreasing because the two branes are approaching one another and
the fifth dimension is collapsing.

 In the M-theory models which we consider in this paper,
    we shall simply impose
   a constraint on moduli space which ensures that the
   distance between the visible and hidden branes becomes fixed
   after collision, as required to converge to the Ho\v rava-Witten
   picture. The constraint 
   corresponds to fixing $a_0/a_1$ after collision,
   forcing a discontinuity in both $\dot{a}_1$ 
   and $\dot{a}_0$.  In this
   case the momentum matching condition
   (continuity of $\dot{a}$) yields rather
   paradoxical behavior in which both $\dot{a}_1$ and $\dot{a}_0$
   reverse after collision, so the universe collapses.
   This does not seem
   physically
   plausible, especially when
  matter is produced at collision, and 
   the expansion of the matter would then have to be reversed
   as the size of the fifth dimension became fixed. More plausible
   	is that $\dot{a}$ is also discontinuous:
   the branes collide, their separation becomes fixed,
   and the pair continue in the same direction of motion (expansion)
   as before collision. Here we simply wish to flag this issue
   as one that we have not resolved in the M-theory models
   considered here: more work is needed to do so.

   We have identified at least one mechanism
   for avoiding contraction or
   collision while still remaining within the moduli 
   space approximation.
   We have constructed models for branes in AdS
   employing `non-minimal' corrections
   to the kinetic terms of $\a_{1,0}$, which are allowed
   by four dimensional general coordinate invariance.
   These non-minimal kinetic terms
   both stabilize the size of the extra dimension and
   allow final expansion from static initial conditions,
with negative potentials.\cite{future}

   Another possibility is that
   the moduli space approximation break-down at collision
   (it must break down, since radiation is produced) leads
   to the release of radiation into the bulk. This is prohibited by
   planar symmetry in the AdS example, but is possible in the
   more general M-theory context.
   Radiation emitted into the bulk contributes to the
   pressure $T_{55}$, which, from the $G_{55}$ Einstein
   equations, acts to decelerate $a_1$.
   The emitted radiation is redshifted as it 
   crosses the bulk, so is likely
   to have less effect on the hidden brane if it is
   absorbed there. The net result would be a slowing of $a_1$,
   causing
   the effective scale factor $a$ to increase.

We do not want to understate the challenge of obtaining a final
expanding universe with stabilized fifth dimension.
In a conventional four-dimensional theory (Einstein gravity plus scalar
fields) it would simply be impossible
to start from zero energy and, through evolution involving negative
potentials, obtain a final expanding universe. Our point is that brane
world scenarios offer ways around this `no-go theorem', which
we have just begun to explore.

   The  AdS examples we have discussed are instructive in that
   they are easier to analyze than the full M-theory case. 
 However, as mentioned above, it is unlikely that
   these model theories are quantum mechanically consistent.
   The most obvious problem is that fine tuning is needed
   to balance the brane tension against the cosmological term.
   Without this balance, no static solutions are possible.
   Computing the quantum corrections may in fact be
   impossible
   since, in the thin-brane limit, these
   are generally infinite and non-renormalizable.

   Therefore, for the remainder of the paper,
   we turn to analogous examples in heterotic M-theory,
   which is more complex but has other advantages.
   The branes in this theory are BPS states, protected
   from quantum corrections by supersymmetry. Their tensions are fixed
   by exact quantum mechanical symmetries and there is
   no fine tuning problem analogous to that present in the
   Randall-Sundrum models.

\subsection{The Background BPS Solution in Heterotic M-theory} \label{bps}

The five-dimensional effective action of heterotic M-theory was derived in 
Refs.~\ref{lukas1} and~\ref{lukas2}. 
Its field content includes a myriad of moduli, most of which will be assumed frozen in this paper.
We shall, therefore, use a simplified action describing gravity $g_{\gamma\delta}$,
the universal ``breathing'' modulus of the Calabi-Yau three-fold $\phi$, 
a four-form gauge field ${\cal A}_{\gamma\delta\epsilon\zeta}$ 
with field strength ${\cal F}=d{\cal A}$ and a single bulk M5-brane.
It is given by
\begin{eqnarray}
\nonumber
& & S=\frac{M_5^3}{2}\int_{{\cal M}_5} d^5x\sqrt{-g}\left({\cal R}-\frac{1}{2}(\partial\phi)^2-\frac{3}{2}\frac{e^{2\phi}{\cal F}^2}{5!}\right) \\
& & \;\;\;\;\;\;-3\sum_{i=1}^3\alpha_iM_5^3\int_{{\cal M}_4^{(i)}} d^4\xi_{(i)}\left(\sqrt{-h_{(i)}}e^{-\phi}
  -\frac{\epsilon^{\mu\nu\kappa\lambda}}{4!}{\cal A}_{\gamma\delta\epsilon\zeta}\partial_{\mu}X^{\gamma}_{(i)}\partial_{\nu}X^{\delta}_{(i)}\partial_{\kappa}X^{\epsilon}_{(i)}\partial_{\lambda}X^{\zeta}_{(i)}\right),
\label{eq:5daction}
\end{eqnarray}
where $\gamma, \delta,\epsilon, \zeta =0,\dots,4$, $\mu,\nu,\dots=0,\dots,3$. 
The space-time is a five-dimensional manifold ${\cal M}_5$ with coordinates $x^{\gamma}$. 
The four-dimensional manifolds ${\cal M}_4^{(i)}$, $i=1,2,3$ are the visible, 
hidden, and bulk branes respectively, and have 
internal coordinates $\xi^{\mu}_{(i)}$ and tension $\alpha_iM_5^3$. 
Note that $\alpha_i$ has dimension of mass. If we denote 
$\alpha_1\equiv-\alpha$, $\alpha_2\equiv\alpha-\beta$, and 
$\alpha_3\equiv\beta$, then the visible brane has tension 
$-\alpha M_5^3$, the hidden brane $(\alpha-\beta) M_5^3$, and the bulk brane
$\beta M_5^3$.
It is straightforward to show that the tension of the bulk brane, $\beta M_5^3$, 
must always be positive. Furthermore, one can easily deduce that the tension on the visible
brane, $-\alpha M_5^3$, can be either positive or negative.
The ekpyrotic scenario can be applied, in principle, to any such vacua.
In this paper, for specificity, we will always take $\alpha>0$, so that the tension on the visible brane 
is negative. Furthermore, we will choose $\beta$ such that $\alpha-\beta>0$, that is, the tension of the 
hidden brane is positive.
The tensor $h^{(i)}_{\mu\nu}$ is the induced metric (and $h_{(i)}$ its determinant) on ${\cal M}_4^{(i)}$.
The functions $X^{\gamma}_{(i)}(\xi_{(i)}^{\mu})$ are the coordinates in ${\cal M}_5$ of a point on ${\cal M}_4^{(i)}$ with coordinates $\xi_{(i)}^{\mu}$.
In other words, $X^{\gamma}_{(i)}(\xi_{(i)}^{\mu})$
describe the embedding of the branes into ${\cal M}_5$.

The BPS solution of Lukas, Ovrut, and Waldram~\cite{lukas1} is then given by\footnote{We have changed the notation used in Ref.~\ref{lukas1} by replacing their $H(y)$ with $\D(y)$. In this paper, we will use the symbol $H$ to denote the Hubble 
parameter. Furthermore, comparing with their notation, we have rescaled $\alpha$ by a factor of $\sqrt{2}/3$ and have defined $e^{\phi}=V$.} 
\begin{eqnarray}
\nonumber
& & ds^2=\D(y)(-N^2d\t^2+A^2d\vec{x}^2)+B^2\D^4(y)dy^2 \\
\nonumber
& & e^{\phi}=B\D^3(y) \\
\nonumber
& & {\cal F}_{0123Y}=-\alpha A^3NB^{-1}\D^{-2}(y)\;\;\;\;\;\;\;\;\;\;\;\;\;{\rm for}\;\;y<Y \\
& & \;\;\;\;\;\;\;\;\;\;\;\;=-(\alpha -\beta)A^3NB^{-1}\D^{-2}(y)\;\;\;\;{\rm for}\;\;y>Y, 
\label{eq:static}
\end{eqnarray}
where 
\begin{eqnarray}
\nonumber
& & \D(y)=\alpha y +C \;\;\;\;\;\;\;\;\;\;\;\;\;\;\;\;\;\;\;\;\;\;\;\;\;\;{\rm for}\;\;y<Y \\
& & \;\;\;\;\;\;\;\;\;=(\alpha-\beta)y+C+\beta Y \;\;\;\;\;\;\;{\rm for}\;\;y>Y,
\label{ansatz}
\end{eqnarray}
and $A,B,C,N$ and $Y$ are constants. 
Note that $A,B,C,N$ are dimensionless and
$Y$ has the dimension of length.
The visible and hidden boundary branes are located at $y=0$ and $y=R$, respectively, and
the bulk brane is located at $y=Y$, $0\leq Y\leq R$.
We assume that $C>0$ so that the curvature singularity at $\D=0$ does not fall between the boundary branes.
Note that $y=0$ lies in the region of smaller volume while $y=R$ lies in the region of larger volume.

Finally, note that inserting the solution of the four-form equation of motion
into Eq.~(\ref{eq:5daction}) yields precisely the bulk action 
given in Ref.~\ref{lukas1} with charge
$-\alpha$ in the interval $0 \leq y \leq Y$ and charge $-\alpha+\beta$ in the
interval $Y \leq y \leq R$. The formulation of the action 
 Eq.~(\ref{eq:5daction}) 
using the
four-form ${\cal{A}}$ is particularly useful when the theory contains bulk
branes, as is the case in ekpyrotic theory.

\subsection{The Moduli Space Action of Heterotic M-theory} \label{modaction}

As in Section~\ref{moduli},
we shall use the moduli space approximation to study the dynamics of 
heterotic M-theory with a bulk brane.  
The static BPS solution involves five constants $A,B,C,N$, and $Y$. These now become the moduli fields, 
$Q^I = (A(\vec{x},\t),
B(\vec{x},\t), \, C(\vec{x},\t), \, N(\vec{x},\t), \, Y(\vec{x},\t))$.
In the limit of homogeneity and isotropy,
the moduli fields  are functions of time only. 
Substituting the static ansatz~(\ref{eq:static}) 
into the action~(\ref{eq:5daction}), and integrating over $y$, 
we obtain the moduli space action ${\cal S}_{mod}$ with 
Lagrangian density
\begin{equation}
{\cal L}_{mod}=G_{IJ}(Q)\dot{Q}^I\dot{Q}^J-V(Q)= {\cal L}_{bulk}+{\cal L}_{\beta},
\end{equation}
where
\begin{eqnarray}
\nonumber
& & {\cal L}_{bulk}=-\frac{3A^3BI_{3}M_5^3}{N}\left\{\left(\frac{\dot{A}}{A}\right)^2+\left(\frac{\dot{A}}{A}\right)\left[\left(\frac{\dot{B}}{B}\right)+\frac{3I_2}{I_3}\dot{C}+\frac{3I_{2b}}{I_3}\beta\dot{Y}\right]\right\} \\
\nonumber  
& & \;\;\;\;\;\;\;\;\;\;\;\; -\frac{3A^3BI_{3}M_5^3}{N}\left\{-\frac{1}{12}\left(\frac{\dot{B}}{B}\right)^2+\frac{I_1}{2I_3}\dot{C}^2+\frac{I_{1b}}{I_3}\dot{C}\beta\dot{Y}+\frac{I_{1b}
}{2I_3}\beta^2\dot{Y}^2\right\} \\
& & {\cal L}_{\beta}=\frac{3\beta M_5^3
A^3B}{N}\left[\frac{1}{2}\D^2(Y)\dot{Y}^2-N^2V(Y)\right]
\label{eq:4daction}
\end{eqnarray}
and 
\begin{eqnarray}
\nonumber
& & I_{ma}\equiv 2\int_{0}^{Y}\D^mdy=\frac{2}{\alpha(m+1)}[(\alpha Y+C)^{m+1}-C^{m+1}] \\
\nonumber
& & I_{mb}\equiv 2\int_{Y}^{R}\D^mdy=\frac{2}{(\alpha-\beta)(m+1)}[((\alpha-\beta)R+C+\beta Y)^{m+1}-(\alpha Y+C)^{m+1}] \\
& & I_{m}\equiv I_{ma}+I_{mb}.
\label{eq:I}
\end{eqnarray}
Note that $I_m$ has the dimension of length.
We see from Eq.~(\ref{eq:4daction}) that the Lagrangian of the 4d effective theory is the sum of two parts, ${\cal L}_{bulk}$ and ${\cal L}_{\beta}$.
The first contribution, ${\cal L}_{bulk}$, comes from the bulk part of the 
five-dimensional action, whereas the second contribution, ${\cal L}_{\beta}$, is the Lagrangian of the bulk brane.
Note that we have added by hand a potential $V(Y)$ in ${\cal L}_{\beta}$ 
which is meant to describe non-perturbative interactions between the bulk 
brane and the boundary branes\cite{moore,lima,derrin}. 
The actions of the two boundary branes, which are at fixed values of $y$, 
do not contribute to the 4d effective action.
Their contribution is canceled by bulk terms upon integration over $y$.
This cancellation is crucial, since it yields a 4d 
effective theory with no potentials for $A,B,C$, or $N$,
thereby confirming that these fields are truly moduli of the theory.

Eq.~(\ref{eq:4daction}) is analogous to the action for gravity with scale factor $A$ coupled to scalar fields. 
Since the overall factor of $BI_3M_5^3$ will generically 
be time-dependent before collision, it follows that the scalar fields $B,C$, and $Y$ are non-minimally coupled to gravity. 
In order to match onto a theory with fixed Newton's constant $G_4$, 
we shall impose the condition
that this factor become constant after collision. 
Hence, after collision, we can identify the 4d effective Planck mass as
\begin{equation}
M_{pl}^2=\frac{BM_5^2(I_3M_5)}{\alpha R+C}
\label{eq:mplA}
\end{equation}
where $I_{3}$ is evaluated at $Y=0$ and for $B,C$ at the moment of collision.
Note that in the limit $(\alpha-\beta) R\ll C$, this expression agrees with the 4d Planck mass identified in Ref.~\ref{witten2}.

At this point, one can define a new scale factor  $\a \equiv
A (BI_3M_5)^{1/2}$, as well as $\n \equiv N (BI_3M_5)^{1/2}$. 
This has the effect of
removing off-diagonal terms in the moduli space metric that couple the
field $A$ to the other variables. 
In these new variables, the bulk Lagrangian becomes 
\begin{eqnarray}
\nonumber
& & {\cal L}_{bulk}=\frac{3\a^3M_5^2}{\n}
\left\{-\left(\frac{\dot{\a}}{\a}\right)^2
+\frac{1}{3}\left(\frac{\dot{B}}{B}\right)^2
+\frac{1}{2}\frac{\dot{B}}{B}\frac{\dot{I}_3}{I_3}\right\} \\
& & \;\;\;\;\;\;\;\;\;\; +\frac{3\a^3M_5^2}{\n}\left\{
\left(\frac{9I_2^2}{4I_3^2}-\frac{I_1}{2I_3}\right)\dot{C}^2
+\left(\frac{9I_2I_{2b}}{2I_3^2}-\frac{I_{1b}}{I_3}\right)\dot{C}\beta\dot{Y}
+\left(\frac{9I_{2b}^2}{4I_3^2}-\frac{I_{1b}}{2I_3}\right)
\beta^2\dot{Y}^2\right\}. 
\label{eq:4dactionb}
\end{eqnarray}
In this form, it is clear that $\a$ is an  scale factor 
analogous to the variable $a$ of the previous section.
The moduli $B$, $C$ and $Y$ behave effectively as scalar fields, 
albeit with nontrivial kinetic terms.

In analogy with the example in Section~IV A and
the discussion following  Eq.~(\ref{eq:m13b}),
we now impose two 
constraints consistent with four dimensional covariance.
These reduce the moduli degrees of freedom and simplify the system.
Namely, we shall impose that 
\begin{eqnarray}
\nonumber
& & B=constant \\
& & C=constant.
\label{eq:conditions}
\end{eqnarray}
At the moment of collision, the modulus $Y$ disappears from 
the theory. The above conditions then imply that the distance between 
the boundary branes as well as the volume of the Calabi-Yau three-fold become fixed.
This is necessary if we want to match onto a theory with fixed gravitational and gauge coupling constants.
For instance, since $I_3$ becomes constant at collision, it follows from Eq.~(\ref{eq:mplA}) that $M_{pl}$ freezes at that point.

If we impose Eq.~(\ref{eq:conditions}), and if we further assume that 
the tension of the bulk brane is small compared to that of the boundary branes, that is, $\beta\ll \alpha$ (which allows us to neglect the correction to the kinetic term for $Y$ coming from ${\cal L}_{bulk}$), then the full Lagrangian reduces to
\begin{equation}
{\cal L}=\frac{3M_5^2\a^3}{\n}\left\{-\left(\frac{\dot{\a}}{\a}\right)^2
+\frac{\beta}{I_3}\left[\frac{1}{2}D(Y)^2\dot{Y}^2-\n^2\frac{V(Y)}{BI_3M_5}\right]\right\}.
\label{eq:final4d}
\end{equation}
This action describes a scalar field $Y$ minimally coupled to
a gravitational background with scale factor $\a$.
Note that the gravitational coupling constant associated with $\a$ is simply $M_5$.

A repeating theme of this paper 
is that the effective scale factor
(and the associated Hubble parameter) can 
be defined several different ways
involving different combinations of 
moduli fields, depending on the physical question being addressed.
For example, looking ahead,
we will show that $\a$ is not the same as the
effective scale factor for an observer on a hypersurface of constant $y$ or the 
effective 
scale factor relevant for describing fluctuations in $Y$.

\subsection{Equations of Motion, the Ekpyrotic Temperature, and
the Horizon Problem} \label{eom}

From the moduli space action obtained in Section~\ref{modaction},
we can find the equation of motion for the bulk brane,
described by $Y$,
in the limit that the bulk brane tension is small, $\beta \ll \alpha$.  
We will use this equation to compute the ekpyrotic temperature, 
the temperature immediately after the branes collide and the 
universe bursts into the big bang phase.  We will then compare the
Hubble radius at the beginning of the big bang phase to the 
causal horizon distance estimated by computing the time it takes
for the bulk brane to traverse the fifth dimension.

Recall that for the static BPS solution, $A$ and $N$ are constants. 
Without loss of generality, we can choose  $A=N=1$ in the BPS limit.
Variation of Eq.~(\ref{eq:final4d}) with respect to $\n$ yields the Friedmann equation which,  setting $\n=\a$, is given by
\begin{equation}
\H^2\equiv \left(\frac{\dot{\a}}{\a^2}\right)^2=\frac{\beta M_5}
{B(I_3M_5)^2}\left(\frac{1}{2}D(Y)^2\dot{Y}^2+V(Y)\right).
\label{eq:N}
\end{equation}
Here we have introduced $\H$ to denote the Hubble constant associated with $\a$. 
Since the gravitational coupling constant associated with $\a$ is $M_5$, we can identify from Eq.~(\ref{eq:N}) the energy density of the bulk brane
\begin{equation}
\rho_{\beta}=3M_5^2\H^2.
\label{eq:rhob}
\end{equation}
The equation of motion for $\a$ yields the second equation of FRW cosmology
\begin{equation}
\frac{\ddot{\a}}{\a}\approx -\frac{\beta}{I_3}\left(\frac{1}{2}D(Y)^2\dot{Y}^2-2V(Y)\right).
\label{eq:ddotaa}
\end{equation}
Finally, we can express the equation of motion for $Y$ 
in a simple way by defining $\Psi$ such that $\dot{\Psi}=(D(Y)/(I_3M_5)^{1/2})\dot{Y}$.
Once again making the gauge choice $n=\a$, one finds that $\Psi$ satisfies
\begin{equation}
\frac{d}{d\tau}\left(\frac{1}{2}\a^{-2}\dot{\Psi}^2+V_{eff}(\Psi)\right)=-3\left(\frac{\dot{\a}}{\a^3}\right)\dot{\Psi}^2,
\label{eq:psi}
\end{equation}
where $V_{eff}\equiv V(Y)/(B(I_3M_5)^2)$.
In this form, Eq.~(\ref{eq:psi}) looks like the equation of motion 
for a scalar field in a cosmological background. It is, therefore, simple to analyze.
The left hand side is the time derivative of the total energy
density associated
with the motion of $Y$.  The right hand side either 
decreases or increases due to the cosmic evolution described by the scale
factor $\a$.  Hence, as regards the kinetic energy of the bulk 
brane, $\a$ is the relevant scale factor.

Perturbing around the BPS limit, we have
$A=1+{\cal O}(\beta/\alpha)$ and
 $I_3=I_3^{(0)}+{\cal O}(\beta/\alpha)$, where $I_3^{(0)}$ is the value of $I_3$ when $\beta=0$ and is time-independent. 
Therefore, $\a=(BI_3^{(0)}M_5)^{1/2}+{\cal O}(\beta/\alpha)$.
If $V(Y)\leq 0$, then all contributions on the   right hand side
of Eq.~(\ref{eq:ddotaa}) cause $\ddot{\a}$ to be negative.
Hence, beginning from a static initial condition ($\dot{\a}=0$), 
$\a$ will contract.
If $\a$ contracts, then, from Eq.~(\ref{eq:psi}), the total energy of $\Psi$ grows with time.
Thus, as long as $V$ is negative semi-definite, the field $\Psi$ gains energy as the bulk brane proceeds through the extra dimension.

At the moment of collision, the modulus $Y$ effectively goes away and 
new moduli describing the small instanton transition and new vector bundle take its place.
In particular, the potential for $Y$ matches onto the potential for these new moduli.
While $V(Y)$ will generically be negative up to the 
moment of collision, we shall assume that, once the other moduli 
are excited during the small instanton transition,
the potential rises back up to zero in the 
internal space of the new moduli such as to leave the 
universe with vanishing cosmological constant.
We base this assumption on the notion that
both the initial and final states consist of two branes in a 
BPS vacuum with $V=0$.
That is,  if we were to adiabatically
detach the bulk brane from the hidden brane
and, then,  transport and attach it 
 to the visible brane, both the initial and final states would 
 consist of two branes only in a BPS configuration with 
 cosmological constant zero.

To mimic the effect of the other moduli during the small instanton 
transition, we shall assume that $V(Y)$ is negative and
approaches zero as $Y \rightarrow 0$.
Note that if $\a$ were constant, the kinetic energy and 
the total energy, $\rho_{\beta}$, would go to zero as $V \rightarrow 0$ and
as the branes collide. However, 
 from Eq.~(\ref{eq:psi}), we see that $\Psi$ (and 
therefore $Y$) has extra kinetic energy as
 $V\rightarrow 0$ due to the gravitational blue shift
effect caused by a contracting scale factor $\a$.
We assume that the extra kinetic energy
(equal to $\rho_{\beta}$ at collision)
is converted into excitations of light degrees of freedom, 
at which point the radiation-dominated era begins.
The temperature after the conflagration that arises from the 
brane collision is referred to as the ``ekpyrotic temperature,"
analogous to the reheat temperature after inflation.

Conceivably, some fraction of the extra kinetic energy
is converted into thermal excitations, 
some into coherent motion of moduli fields, and some, perhaps,
into bulk excitations (gravitons).  
If coherent motion is associated with massless degrees of freedom,  
the associated kinetic 
energy redshifts away  
faster than radiation. If associated degrees of freedom
are massive, they can ultimately decay into radiation. 
Neither case is problematic.  For simplicity,
we shall just assume that the kinetic energy at 
collision is converted entirely into radiation with an efficiency of order unity.

The collision energy  can be computed by integrating
the equation of motion,
Eq.~(\ref{eq:psi}), which is expressed explicitly in terms of the
time derivative of the total bulk brane energy.
However, it is  somewhat
simpler to first compute  the Hubble parameter $\H_c$  
upon collision using Eq.~(\ref{eq:ddotaa}), and then to substitute in Eq.~(\ref{eq:rhob}) to obtain the collision energy $\rho_{\beta}$.
(Note that the subscript $c$ denotes that the quantity is evaluated at the moment of collision.)
Eq.~(\ref{eq:ddotaa}) can be rewritten as
\begin{equation}
\frac{\ddot{\a}}{\a}\approx -\frac{\beta}{I_3}\left(\frac{1}{2}\D(Y)^2\dot{Y}^2+V(Y)-3V(Y)\right)\approx -\frac{\beta}{I_3}(-3V(Y)),
\label{eq:ddota2}
\end{equation}
where we have used the fact that, to leading order in $\beta/\alpha$,  
\begin{equation}
\frac{1}{2}\D^2(Y)\dot{Y}^2+V(Y)=E \approx 0.
\label{eq:eomm}
\end{equation}
Note that $E$ is the total energy of the bulk brane which is assumed small compared to the energy gained from gravity.
We can then integrate Eq.~(\ref{eq:ddota2}) to obtain
\begin{equation}
\frac{\dot{\a}}{\a}=-3\int_{Y=R}^{Y=0}\frac{\beta}{I_3}(-V(Y))\frac{d\t}{dY}dY=-\frac{3\beta}{\sqrt{2}I_3}\int_{Y=0}^{Y=R}D(-V)^{1/2}dY,
\label{eq:ddota3}
\end{equation}
where we have made use of the fact that $I_3=I_3^{(0)}=constant$ to leading order in $\beta/\alpha$.
Since $\a\approx (BI_3^{(0)}M_5)^{1/2}$, we obtain
\begin{equation}
\H_c=\left\vert\frac{\dot{\a}}{\a^2}\right\vert=\frac{3\beta M_5}{\sqrt{2B}(I_3M_5)^{3/2}}\int_{Y=0}^{Y=R}D(-V)^{1/2}dY.
\label{eq:h}
\end{equation}
Now that we have an expression for the Hubble parameter $\H_{c}$, we can substitute in Eq.~(\ref{eq:rhob}) and find
\begin{equation}
\rho_{\beta}=\frac{27\beta^2M_5^4}{2B(I_3M_5)^3}\left(\int_{Y=0}^{Y=R}D(-V)^{1/2}dY\right)^2.
\end{equation}
The corresponding ekpyrotic temperature is then
\begin{equation}
\frac{T}{M_{pl}}\sim \left(\frac{\rho_{\beta}}{M_5^4}\right)^{1/4}=\frac{3^{3/4}\sqrt{\beta}}{(2B)^{1/4}(I_3M_5)^{3/4}}\left[\int_{Y=0}^{Y=R}D(-V)^{1/2}dY\right]^{1/2}.
\label{eq:T}
\end{equation}
(N.B. The identification of energy density or effective Planck mass
may vary under Weyl transformation, but 
the ratio $T/M_{pl}$ is invariant.)    
For instance, consider a potential of the form $V(Y)=-ve^{-\m\alpha Y}$, 
where $v$ and $m$ are positive, dimensionless constants.
Since non-perturbative potentials derived from string and M-theory are 
generically of exponential form (for motivation, see the discussion under Eq.~(\ref{eq:expo})), 
this potential will be a standard example throughout.
In that case, the temperature is calculated, using Eq.~(\ref{eq:T}), to be 
\begin{equation}
\frac{T}{M_{pl}}\approx \frac{3^{3/4}(2v)^{1/4}}{(I_3M_5)^{1/2}(\alpha R +C)^{1/4}}\left(\frac{M_5}{M_{pl}}\right)^{1/2}\left(\frac{\beta}{\alpha}\right)^{1/2}\frac{(\m C+2)^{1/2}}{\m}.
\label{eq:Texpo}
\end{equation}
where we have used Eq.~(\ref{eq:mplA}).
As an example, we might suppose $\alpha=2000M_5$, $\beta=M_5$, $B=10^{-4}$, $C=1000$, $R=M_5^{-1}$, $v\sim 10^{-8}$, and $\m=0.1$, all plausible values.
This gives $M_5=10^{-2}M_{pl}$ and produces an ekpyrotic temperature of $10^{11}$~GeV.  
Note that, with these parameters, the magnitude of the potential energy density for $Y$ 
is $(10^{-6}M_{pl})^4$ at collision.
Thus, the typical energy scale for the potential is $10^{13}$~GeV.
 Later, we will see that these same parameters produce 
an acceptable fluctuation amplitude. 
We want to emphasize, however, that there is a very wide range of parameters 
that lead to acceptable cosmological scenarios. 

Let us now turn our attention to the homogeneity problem. 
We have argued that the universe begins in a BPS state,
which is homogeneous.  This condition is stronger than needed to
solve the homogeneity problem.
It suffices that the universe be homogeneous on scales smaller than
the causal, particle horizon. 
Let $(-\t)_{tot}$ denote the time taken by the bulk brane to
travel from the hidden to the visible brane. 
By integrating Eq.~(\ref{eq:eomm}), we find that the comoving time is
\begin{equation}
(-\t)_{tot}=\int_{0}^{R}\frac{\D(Y') dY'}{\sqrt{-2V(Y')}}.
\label{eq:tt}
\end{equation}
The horizon distance $d_{HOR}$, as measured by 
an observer on the visible brane, is the elapsed comoving
time at collision ($Y=0$) times the scale factor, $D^{1/2}(y=0) =C^{1/2}$. We
find, for an exponential potential of the above form, that
\begin{equation}
d_{HOR}=C^{1/2}(-\t)_{tot}\approx \frac{\sqrt{2C}}{\m\alpha\sqrt{v}}(\alpha R+C)e^{\m\alpha R/2}.
\end{equation}
On the other hand, the Hubble radius at collision is obtained from Eq.~(\ref{eq:h})
\begin{equation}
\H_c^{-1}=\frac{m^2I_3(BI_3M_5)^{1/2}}{3\sqrt{2v}(mC+2)}\left(\frac{\alpha}{\beta}\right).
\end{equation}
The causal horizon problem is solved if the particle horizon at collision satisfies
\begin{equation}
\frac{d_{HOR}}{\H_c^{-1}}\sim e^{m\alpha R/2} > \left(\frac{T}{M_{pl}}\right)\cdot e^{70}.
\label{eq:dhor}
\end{equation}
The condition  is easily satisfied for the values of parameters mentioned 
above, where $m \alpha R/2 \sim 10^2$, the equivalent of 100 e-folds of 
hyperexpansion in an inflationary model.

\subsection{Cosmological Evolution for an  Observer at Fixed y}

We have seen that  the scale factor relevant to describing
the equation of motion for $Y$ is
 $a$, a particular combination of moduli fields. 
 Furthermore, if we assume nearly BPS initial conditions (that is, vanishing potential and kinetic energy), then $a$ is a decreasing function of time.
Hence, the scalar field $Y$ evolves as if the universe is contracting.

However, as pointed out previously, an important feature 
is that the effective scale factor 
for other physical quantities depends on other combinations of moduli fields.
In this subsection, we shall derive the cosmological evolution as seen 
by an observer living on a hypersurface of constant $y$ (for example, an observer on the visible brane). 
We will find that the scale factor seen by such family of observers is different than $a$. 
In particular, we find that any such observer sees the universe {\it expanding} before the bulk brane collides with the visible brane.

Consider, for concreteness, an observer living on the visible brane.
The induced metric on that hypersurface is obtained from Eq.~(\ref{eq:static}) 
\begin{equation}
ds^2_{y=0}=\frac{a^2C}{BI_3M_5}(-d\t^2+d\vec{x}^2)\equiv
a_1^2(-d\t^2+d\vec{x}^2).
\end{equation}
The rate of change of the induced scale factor $a_1$ can be written as
\begin{equation}
\frac{\dot{a}_1}{a_1}=\frac{\dot{a}}{a}-3\beta\frac{I_{2b}}{I_3}\dot{Y}=\frac{\dot{a}}{a}+3\beta\frac{I_{2b}}{I_3}\frac{\sqrt{-2V(Y)}}{D(Y)},
\label{eq:a0}
\end{equation}
where we have used Eq.~(\ref{eq:eomm}).
In this way, we have expressed $\dot{a}_1$ as the sum of two contributions: the first contribution is given by $a$ and tends to make $a_1$ contract; the second term, coming from $I_3$, is positive and tends to make $a_1$ expand.
To determine which of these two terms dominates at the moment of collision, 
we note from Eq.~(\ref{eq:ddota3}) that 
\begin{equation}
\frac{\dot{a}}{a}\sim (-V(Y))^{1/2}.
\end{equation}
Therefore, both terms on the right hand side
of Eq.~(\ref{eq:a0}) are proportional to $(-V(Y))^{1/2}$.
However, the coefficients are different functions of $\alpha$ and $C$. 
For the case of the exponential potential, for instance, one finds that 
reasonable values of $\alpha$, $C$, and $\m$ (such as those given at the 
end of Section~\ref{eom}) result in the $I_3$ term being larger than 
the $a$ term in Eq.~(\ref{eq:a0}).
The net effect is to make $a_1$ grow with time;
that is,
an observer at $y=0$ sees the universe expanding.   
This is in agreement with the results for the AdS case presented in Section~\ref{moduli}.

For other hypersurfaces of constant $y$, a similar story holds. 
The induced scale factor is the product of $a$ which decreases and a function of $Y$ which increases.
Once again, reasonable values of the parameters result in expanding 
hypersurfaces at the time of collision.
In particular, $a_0$, the scale factor on the $y=R$ hidden brane, is expanding at collision, in agreement with the AdS results.

To summarize, for shallow potentials (for example, an exponential potential) 
and reasonable values of the parameters, we have seen that both $a_1$ and $a_0$ are expanding at the moment of collision. 
On the other hand, $a$ is contracting.
Since this agrees qualitatively with the AdS case (see Fig. 3), 
the discussion at the end of Section~\ref{moduli} concerning the 
matching condition at collision and the subsequent expansion of the 
universe applies also to heterotic M-theory.

\subsection{Cosmological Evolution for an Observer on the Bulk Brane}

So far, we have adopted the point of view of the low-energy four-dimensional effective action. 
It is sometimes useful, for instance in calculating the fluctuation spectrum, to adopt  the point of view of an observer living on the bulk brane. 
The motion of the bulk brane through the curved space-time induces an FRW evolution on its worldvolume.
In terms of conformal time $\eta$ on the bulk brane, the induced metric is given
to lowest order in $\dot{Y}$ by
\begin{equation}
ds^2_4=\D(Y(\eta))(-d\eta^2+d\vec{x}^2).
\label{eq:induced}
\end{equation}
From the form of the metric in Eq.~(\ref{eq:induced}), we see that the scale factor $a_B(\eta)$ describing the FRW evolution on the brane is simply given by $a_B(\eta)\equiv\D^{1/2}(\eta)$.
Since the bulk brane moves from a region of larger $D$ (location of the hidden brane) to a region of smaller $D$ (location of the visible brane), an observer on the bulk brane sees a contracting universe (as opposed to an observer on the visible brane who
 sees expansion).
Finally, we note that for non-relativistic motion, one has $\eta\approx \t$, where $\t$ is global conformal time. 

\subsection{Summary of homogeneous propagation of the bulk brane}

In this section, we have described the spatially homogeneous propagation of the 
bulk brane in terms of 
the evolution of three different quantities: the scale factor 
as felt by the modulus $Y$ of the bulk brane, the scale factor for an 
observer living on a hypersurface of constant $y$, and the scale factor for an observer on the bulk brane.
For nearly BPS initial conditions, 
we have seen that the scale factor that appears in the equation
of motion for the bulk brane, namely $\a$, 
decreases with time.
This 
 means that there is a gravitational blue shift 
effect that increases the kinetic energy of the bulk brane.
This added energy, we propose,
is converted to radiation and matter upon collision.
Furthermore, we have shown that 
observers at fixed $y$ generically see an expanding universe.
Finally, an observer on the bulk brane sees a contracting universe.
This is a simple geometrical consequence of the fact that the bulk brane travels from a region of smaller curvature to a region of larger curvature.

We should mention that both the modulus $Y$ and observers at fixed $y$ see scale factors which are slowly-varying in time, in fact almost constant.
This is because both variations are due to the back-reaction of the bulk brane 
onto the geometry, an effect which is of order $\beta/\alpha$.
On the other hand, the cosmological evolution felt by an observer on the bulk brane is  faster, although only by a logarithmic factor. 
No superluminal expansion is taking place from the point-of-view of any
observer. Rather, what characterizes the ekpyrotic scenario is that all motion
and expansion is taking place exceedingly slowly for an exceedingly long
period of time.

\section{Spectrum of fluctuations} \label{spec}

In this section, we show how the ekpyrotic scenario can
produce a nearly scale-invariant spectrum of adiabatic, 
gaussian, scalar (energy 
density) perturbations that may account for the anisotropy of the 
cosmic microwave background and seed
large-scale structure formation.
The density perturbations  are caused by ripples in the bulk brane 
which are generated by quantum fluctuations as the brane traverses the bulk.
The ripples  result in 3d spatial
variations in the time of collision and thermalization, and, consequently,
they induce temperature fluctuations in the hot big bang phase.

Because both the ekpyrotic scenario and inflationary cosmology
rely on quantum fluctuations to generate adiabatic perturbations,
the calculational formalism for predicting the perturbation spectrum
and many of the equations are 
remarkably similar.  One  difference is that
 inflation entails  superluminal expansion and the
ekpyrotic scenario does not.
For the ekpyrotic scenario,
the fluctuations are generated 
as the bulk brane moves
slowly through the bulk.  For the examples considered here,
the motion is in  the direction in which the warp factor
is shrinking. Because of the shrinking warp factor, the  Hubble 
radius for an observer on the brane is decreasing.
The effect of a decreasing Hubble radius is
to make the spectrum blue.
In inflation, the Hubble radius is expanding in the 4d space-time, and, consequently, 
the spectrum is typically red.

Here, we give an abbreviated version of the derivation
that emphasizes the similarities and differences from the inflationary
case.
For this purpose, we adapt the ``time-delay" approach
introduced by Guth and Pi for the case of inflation to the 
colliding brane  picture.\cite{GuthPi}  This approach
has the advantage that it is relatively simple and intuitive. 
Experts are aware that this approach is 
inexact and non-rigorous\cite{wang} and,  hence,  
might question the reliability.
The  more cumbersome and less intuitive
gauge invariant approach introduced by Bardeen, Steinhardt and Turner\cite{BST} and 
by Mukhanov\cite{Mukh} (see also Ref.~\ref{Mukh2}) is preferable since it 
is rigorous and applies to a wider range
of models. We have developed  the analogue of the gauge invariant 
approach for the colliding branes, and we find that the 
time-delay approach does match for the cases we consider.
We will present the gauge invariant formalism  for the ekpyrotic
 scenario and a  fully detailed analysis 
 in a separate publication~\cite{pert}.

In the first  subsection, we shall assume that the ripples in the
bulk brane have already been generated and begin our
computation just as the bulk brane
 collides with our visible brane.
The
bulk brane  has position 
$Y(\t, \vec{x}) = Y_0(\t) + \delta Y(\t, \vec{x})$,
where $Y_0(\t)$ is the average position of the brane along the bulk
($y$) direction and $\delta Y$ represents the small ripples.
 Our goal is to adapt the time-delay formalism
to compute how the ripples translate into
density fluctuations in the hot big bang phase.  
In the second subsection, we shall
discuss how  $\delta Y$ is set  by quantum fluctuations
and the general conditions under which the fluctuation spectrum
will be nearly scale-invariant.
Then, we will present specific
models that satisfy the scale-invariant conditions and 
discuss general model-building principles and constraints.
In the final subsection, we will present the computation for the case of tensor 
(gravitational wave) perturbations and show that 
the spectrum is tilted strongly 
towards the blue, a prediction that differs significantly from inflationary 
models. 

\subsection{From brane ripples to density fluctuations}

The fluctuations $\delta Y$ result in variations in the time of 
collision ($\delta \t$) 
that depend on the   position along the 
bulk surface, $\vec{x}$.  In this sense, the bulk brane position $Y$ plays 
a role analogous to the inflaton $\phi$  and the fluctuations $\delta Y$ 
play a role similar to inflaton fluctuations $\delta \phi$.
The time-delay formalism
 applies under the assumption that the
time delay is 
 independent of time when the perturbations are well 
outside the horizon; that is, $\delta \t = \delta \t (\vec{x})$.
The formalism, then,  allows one to compute how $\delta \t(\vec{x})$
converts into a density perturbation amplitude.

In the de Sitter limit, one has $\delta \t  \sim
\delta \phi/\dot{\phi}$, where the fluctuations $\delta \phi$
are 
time-independent and $\dot{\phi}$ is also time-independent.  Hence, the 
assumption of the time-delay formalism is satisfied.  In the ekpyrotic
 scenario, $\delta \t \sim \delta Y/ \dot{Y}$. Note that $\delta Y$
is time-dependent, and so is $\dot{Y}$.  However, under circumstances
to be discussed later in this section, the two have the same time-dependence
and, consequently, $\delta \tau$ is time-independent, as required.
(The time-independent condition is only approximate in both scenarios.
The weakness of the time-delay approach is that it cannot be simply
generalized to the  time-dependent case;
 corrections  must be computed using a
gauge invariant formulation~\cite{wang}.)

We shall assume that the stress-energy tensor after collision is 
that of an ideal fluid with pressure $P$ and energy density $\rho =3 P$
\begin{equation}
T^{\mu\nu}=Pg^{\mu\nu}+(P+\rho)u^{\mu}u^{\nu},
\end{equation}
where $u^{\mu}$ is the velocity of the fluid normalized to $u^2=-1$.

The perturbations can be characterized by the Olson\cite{Olsen}
variables,
$S$ and div~$X$, defined by
\begin{eqnarray}
\nonumber
& & S\equiv -1+3\rho\cdot(h^{\mu\nu}\nabla_{\mu}u_{\nu})^{-2} \\
& & {\rm div}X\equiv\nabla_{\mu}(h^{\mu\nu}\nabla_{\nu}\rho),
\label{eq:initcond}
\end{eqnarray}
where $h^{\mu\nu}\equiv g^{\mu\nu}+u^{\mu}u^{\nu}$.
The calculation then proceeds in two steps.
First, we find the value of $S$ and ${\rm div}X$ at the moment of collision.
Secondly, we calculate the time-evolution of $S$ in a radiation-dominated universe.

If the average collision time is $\t=0$,
then $\t' \equiv  \t- \delta\t(\vec{x})=0$ is the time when collision occurs 
at position $\vec{x}$.
We have
\begin{eqnarray}
\label{eq:olsen}
\nonumber
& & S(\t'=0)=-\frac{2}{3\a_c\H_c}\vec{\partial}^2\delta\tau(\vec{x}) \\
& & {\rm div}X\;(\t'=0)=6\a_c^{-2}\H_c\dot{\H}_c\vec{\partial}^2\delta\tau(\vec{x}).
\end{eqnarray}
It is useful to define a dimensionless time variable $x$ by
\begin{equation}
x=\frac{k}{\sqrt{3}}\left(\t'+\frac{1}{\a_c\H_c}\right).
\end{equation}
When $x<1$ ($x>1$), the mode 
is outside (inside) the sound horizon.
Then, as shown by Olson, 
the density perturbation with wavenumber $k$, $\delta_k$,
is related to the Fourier mode $S_k$ via
\begin{equation}
\frac{1}{2x}\frac{d}{dx}(x^2\delta_k)=S_k,
\label{eq:rel}
\end{equation}
where  $S_k$ satisfies the evolution equation 
\begin{equation}
x^2\frac{d^2S_k}{dx^2}-2x\frac{dS_k}{dx}+(2+x^2)S_k=0.
\end{equation}
The solution of the $S_k$ equation
can be written as
\begin{equation}
S_k=C_1x\sin x+C_2x\cos x.
\label{eq:s}
\end{equation}
To fix the coefficients $C_1$ and $C_2$, we use the 
initial conditions given in Eq.~(\ref{eq:olsen}).
In terms of $x$, the collision time is $x_c=k/(\sqrt{3}\a_c\H_c)$, and 
the conditions at collision read
\begin{eqnarray}
\nonumber
& & S_k=2x_c^2\a_c\H_c\Delta\t(k) \\
& & x_c\frac{dS_k}{dx_c}=2\left(1-\frac{\dot{\H}_c}{\a_c\H_c^2}\right)S_k, 
\end{eqnarray}
where $\Delta\t(k)\equiv k^{3/2}\delta\t(k)/(2\pi)^{3/2}$.
Using these conditions, one finds that the coefficients $C_1$ and $C_2$ are 
given by
\begin{eqnarray}
\nonumber
& & C_1=2\a_c\H_c\left(1-\frac{2\dot{\H}_c}{\a_c\H_c^2}\right)\Delta\t(k) \\
& & C_2=\left(1-\frac{2\dot{\H}_c}{\a_c\H_c^2}\right)^{-
1}\left(\frac{2\dot{\H_c}}{\a_c\H_c^2}\right)x_cC_1
\end{eqnarray}

The modes of interest lie far
outside the horizon at the time of collision, that is, $x_c\ll 1$. 
Thus, when $x\gg 1$ (when the mode comes back inside the 
horizon), the second term on the right hand side  of Eq.~(\ref{eq:s}) 
is suppressed by a factor of $x_c$ and is 
therefore negligible.  (This is the ``decaying'' mode.)
Using this fact in integrating Eq.~(\ref{eq:rel}), one obtains
\begin{equation}
|\delta_k|=4\a_c\H_c\left(1-\frac{2\dot{\H}_c}{\a_c\H_c^2}\right)|\Delta\t(k)|.
\label{eq:del1}
\end{equation}
We see that $\delta_k$ is the product of two factors: the factor $a_c\H_c|\Delta\t(k)|$ accounts for the fact that different regions of space heat up and therefore begin to redshift at different times, while the factor in parentheses depends on $\dot{
\H}_c$ 
and describes how the change in the Hubble parameter during
collision affects the fluctuations.
For inflationary models near the de Sitter limit, $\dot{\H_c}\rightarrow 0$,
and so $\delta_k$ is directly related to the time delay $\delta \t(k)$.
For the  ekpyrotic
 model, the scale factor $\a(\t)$ is of the form $const.+{\rm log}\;\t$ and $\dot{\H}/\a\H^2$ is approximately constant.
Hence, once again, $\delta_k$ is directly related to the time delay.
In both scenarios, Eq.~(\ref{eq:del1}) agrees  with the exact 
gauge invariant calculation of density perturbations
except for small corrections to the prefactor.

For example, consider the exponential potential $V=-ve^{-\m\alpha Y}$.
In that case, Eq.~(\ref{eq:del1}) yields
\begin{equation}
|\delta_k|\approx \frac{4\m^2\alpha\sqrt{2v}}{mC+2}|\Delta\t(k)|,
\label{eq:dpot1}
\end{equation}
where we have used Eq.~(\ref{eq:h}), and where we have assumed that $\dot{\H_c}/(\a_c\H_c^2)\gg 1$, a reasonable approximation for the values of $\alpha$, $C$, and $\m$ of interest. 

\subsection{From quantum fluctuations to brane ripples} \label{scaleinv}

Eq.~(\ref{eq:del1}) expresses the density perturbation in terms of
the time delay at the time of collision, $\Delta \t(k)$.
In this section, we compute the spectrum of
 quantum fluctuations of the brane
$\delta Y_k$ and use the result to compute the time delay,
$\Delta \t(k)$.

\subsubsection{The scalar fluctuation equation}\label{scaleinv0}

For the calculation of quantum fluctuations, it is sufficient to work at the lowest order in $\beta/\alpha$. 
Without loss of generality, we can therefore set $A=N=1$.
In that case, the bulk brane Lagrangian is given by
\begin{equation}
{\cal L}_{\beta}=3\beta M_5^3 B\left[\frac{1}{2}\D(Y)^2\eta^{\mu\nu}\partial_{\mu}Y\partial_{\nu}Y -V(Y)\right].
\end{equation}
Note that this agrees with ${\cal L}_{\beta}$ given in Eq.~(\ref{eq:4daction}) when we set $A=N=1$ and spatial gradients of $Y$ to zero.
Let us first consider the spatially homogeneous motion of the brane which will be described by $Y_0(\t)$.
(The subscript ``0'' emphasizes that we want to think of $Y_0$ as the background motion.)
It is governed by the following equation of motion
\begin{equation}
\frac{1}{2}\D(Y_0)^2\dot{Y_0}^2+V(Y_0)=E,
\label{eq:back}
\end{equation}
where $E$ is a constant. 
Eq.~(\ref{eq:back}) is, of course, simply the statement that the energy $E$ of the bulk brane is conserved to this order in $\beta/\alpha$.
Since we have chosen the visible brane to lie at $y=0$ and the hidden 
universe to lie at $y=R$, we focus on the branch $\dot{Y}<0$ in which 
case the bulk brane moves towards the visible brane.
The solution to Eq.~(\ref{eq:back}) is then given by
\begin{equation}
(-\t)=\int_{0}^{Y_0}\frac{\D(Y') dY'}{\sqrt{2(E-V(Y'))}}
\label{eq:t}
\end{equation}
with $\t\leq 0$, and with the collision occurring at $\t=0$.

Let us now consider fluctuations around the background solution $Y_0(\t)$.
Namely, if $Y=Y_0(\t)+\delta Y(\t,\vec{x})$, with $\delta Y(\t,\vec{x})\ll Y_0(\t)$,
we can expand the action to quadratic order in $\delta Y$
\begin{equation}
{\cal L}_{fluc}\sim \frac{1}{2}\D_0^2[-\delta {\dot Y}^2+(\vec{\partial}(\delta Y))^2]+\left[\alpha^2\D_0^{-2}(V_0-E)-\alpha \D_0^{-1}\frac{dV_0}{dY_0}+\frac{1}{2}\frac{d^2V_0}{dY_0^2}\right](\delta Y)^2.
\label{eq:flucaction}
\end{equation}
where we have used Eq.~(\ref{eq:back}), and where we have introduced $\D_0\equiv \D(Y_0)$ and $V_0\equiv V(Y_0)$ for simplicity. 

The key relation is the fluctuation equation
 as derived from the action~(\ref{eq:flucaction})
\begin{eqnarray}
\nonumber
& & x^2\frac{d^2f_{\vec{k}}}{dx^2}-\left[\frac{\ddot{a}_{pert}}{a_{pert}}\t^2-
x^2\right]f_{\vec{k}}=0 \\
& & x\equiv|\vec{k}|(-\t),
\label{eq:f1}
\end{eqnarray}
where $f_{\vec{k}}\equiv \D_0\cdot\delta Y_{\vec{k}}$ and 
where $a_{pert}$ is defined by
\begin{equation}
\frac{\ddot{a}_{pert}}{a_{pert}}\equiv \D_0^{-3}\left(\alpha\frac{dV_0}{dY_0}-D_0\frac{d^2V_0}{dY_0^2}\right).
\label{eq:aeff}
\end{equation}

The fluctuation equation, Eq.~(\ref{eq:f1}), can
be compared with the corresponding equation for the perturbations of a scalar field with no potential and minimally coupled to an FRW background with scale factor $a(\t)$
\begin{equation}
\delta\ddot{\phi}_{\vec{k}}+2\frac{\dot{a}}{a}\delta\dot{\phi}_{\vec{k}}
+k^2\delta\phi_{\vec{k}}=0.
\label{eq:dphi}
\end{equation}
Defining $f_{\vec{k}}=a\cdot\delta\phi_{\vec{k}}$, Eq.~(\ref{eq:dphi}) 
becomes
\begin{equation}
x^2\frac{d^2f_{\vec{k}}}{dx^2}-\left[\frac{\ddot{a}}{a}\t^2-
x^2\right]f_{\vec{k}}=0.
\label{eq:f1a}
\end{equation}

Comparing Eqs.~(\ref{eq:f1}) and~(\ref{eq:f1a}), one sees
that $a_{pert}$ plays the role of an effective background for the 
perturbations.
An observer on the bulk brane sees a scale factor $a_B=\D^{1/2}$ (Eq.~(\ref{eq:induced})) but the fluctuations 
evolve, according to Eq.~(\ref{eq:f1}), as if the scale factor were 
$a_{pert}$.
Hence, the shape of the fluctuation spectrum depends on $a_{pert}$, not 
$a_B=\D^{1/2}$; but the physical wavelength is determined by $a_B=\D^{1/2}$, not 
$a_{pert}$.
This is an important subtlety in our calculation.
Let us now discuss the Hubble horizon for the perturbations. 
Recall that in usual 4d cosmology (see Eq.~(\ref{eq:f1a})), we have 

\begin{equation}
x=k(-\t)=\left(\frac{k}{a}\right)\cdot a\cdot(-\t)=k_{phys}a\cdot(-\t)\sim k_{phys}H^{-1},
\label{eq:hor}
\end{equation}
where $H^{-1}\equiv a^2/\dot{a}$ is the Hubble radius as derived from the
scale-factor $a$. 
By definition, a mode is said to be outside the Hubble horizon when its 
wavelength is larger than the Hubble radius. 
From Eq.~(\ref{eq:hor}), we see that this occurs when $x<1$.
Therefore, a mode with amplitude $f_{\vec{k}}$ crosses outside the horizon when 
$x\sim {\cal O}(1)$.
Similarly, in our scenario we can 
write
\begin{equation}
x=k(-\t)=k_{phys}\D_0^{1/2}(-\t)\equiv k_{phys}H_{pert}^{-1},
\end{equation}
where $k_{phys}=k/\D_0^{1/2}$ (since $a_B=\D_0^{1/2}$ relates comoving 
scales to physical length scales on the bulk brane). 
The role of the Hubble radius is replaced by 
\begin{equation}
H_{pert}^{-1}\equiv \D_0^{1/2}(-\t)=\D_0^{1/2}\int_{0}^{Y_0}\frac{\D(Y) dY}{\sqrt{2(E-V(Y))}},
\label{eq:reff}
\end{equation}
which is to be thought of as an effective Hubble radius for the perturbations.
So, as suggested above, the length scale at which
amplitudes freeze depends on $a_B$ (rather than $a_{pert}$), but the 
amplitude itself, as derived from Eq.~(\ref{eq:f1}),
depends on $a_{pert}$.  The feature  of two different scale factors
is a novel aspect of the ekpyrotic scenario.

By the time the bulk brane collides with the visible brane, modes are frozen on all scales less than the value of $H_{pert}^{-1}$ when the bulk brane leaves the hidden brane.
Comparing Eqs.~(\ref{eq:dhor}) and~(\ref{eq:reff}), we see that this initial value of $H_{pert}^{-1}$ is of the order of the particle horizon $d_{HOR}$ at the moment of collision. 
Recall from Section~\ref{eom} that $d_{HOR}$ is required to be exponentially larger than the Hubble radius at collision, $\H_c^{-1}$, in order to solve the homogeneity problem.
Hence, we see that modes are frozen on scales exponentially larger than $\H_c^{-1}$, thereby solving the inhomogeneity problem.

\begin{figure}
{\par\centering \resizebox*{4in}{4in}{\includegraphics{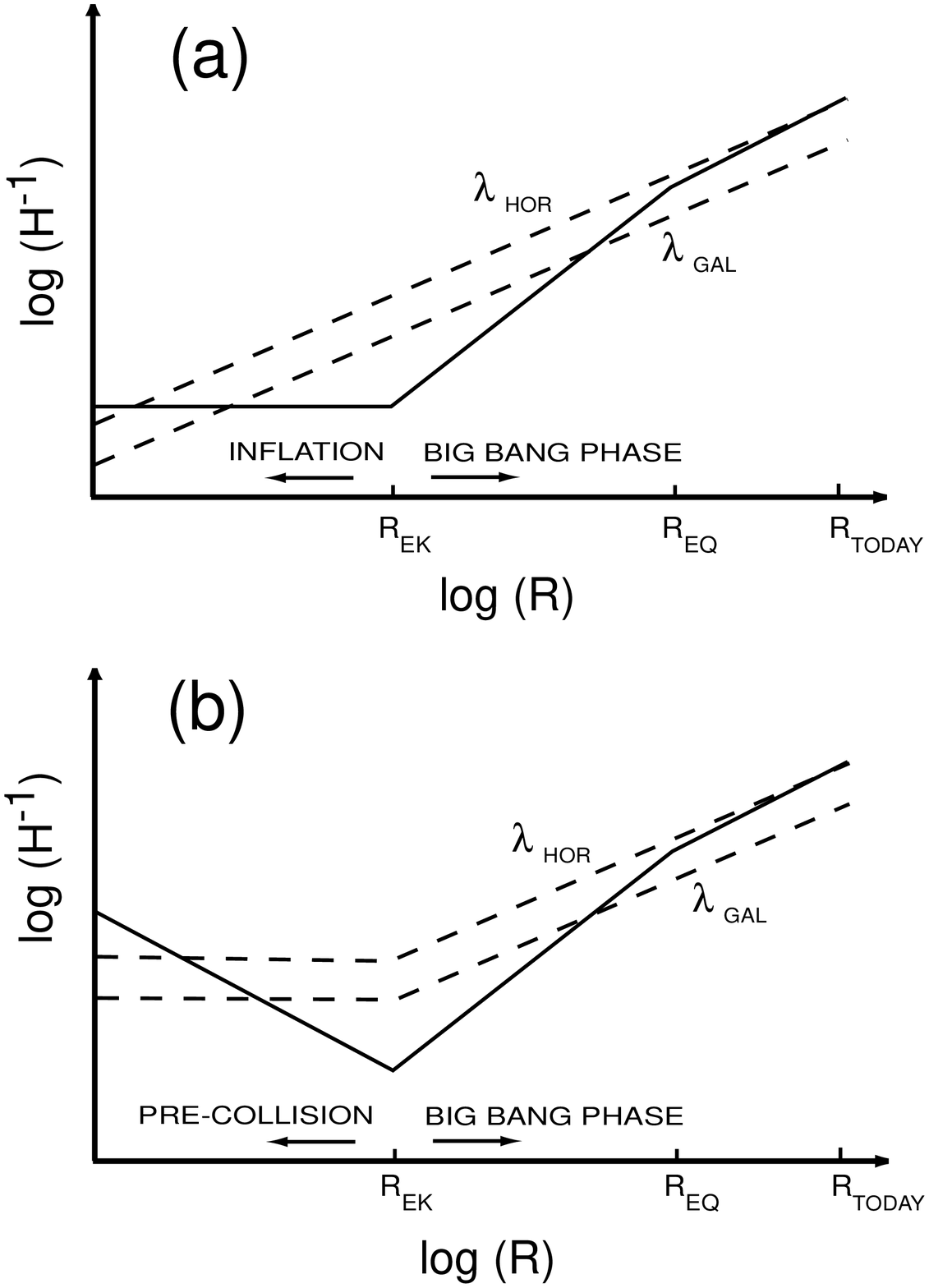}} \par}
 \caption{Sketch comparing the generation of a super-horizon
 spectrum of perturbations in (a) inflationary cosmology  versus
 (b) the ekpyrotic universe.  During inflation, the Hubble radius
 is nearly fixed and the fluctuation wavelength  grows exponentially
 fast, causing modes to be stretched outside the horizon.
 In the ekpyrotic scenario, modes correspond to ripples on the
 moving bulk brane. The perturbations have nearly constant wavelength
 but the effective Hubble radius shrinks, once again causing modes to cross outside the horizon.  }
 \label{fig:comp}
 \end{figure}

The comparison to inflationary cosmology
is made in Figure~\ref{fig:comp}. The salient feature of both models
is that perturbation modes inside the Hubble horizon escape outside
in the early universe and re-enter  much later.  
However, the behavior of the scale factor and the Hubble horizon 
are quite different.  In inflation, the wavelengths are stretched
superluminally while the horizon is nearly constant. In the ekpyrotic
scenario, the wavelengths are nearly constant while the 
horizon shrinks.

It remains to show that we can obtain a spectrum which is scale-invariant.
Writing the equation for the perturbations in the form of Eq.~(\ref{eq:f1}) is useful since one can read off from it the spectral slope of the power spectrum. 
It is determined by the value of $(\ddot{a}_{pert}/a_{pert})\t^2$.
In particular, one obtains a scale-invariant spectrum if $(\ddot{a}_{pert}/a_{pert})\t^2=2$ when the modes observed on the CMB cross outside the horizon.
(Note that in usual 4d cosmology, this is 
achieved for an expanding de Sitter universe with $a \propto -\tau^{-1}$
or a contracting matter-dominated universe with $a \propto \tau^2$.)
Combining Eqs.~(\ref{eq:t}) and~(\ref{eq:aeff}), we find
\begin{equation}
\frac{\ddot{a}_{pert}}{a_{pert}}\t^2=\D_0^{-3}\left(\alpha\frac{dV_0}{dY_0}-D_0\frac{d^2V_0}{dY_0^2}\right)\left[\int_{0}^{Y}\frac{\D(Y') dY'}{\sqrt{2(E-V(Y'))}}\right]^2.
\label{eq:2}
\end{equation}
The spectrum will be scale-invariant if the right hand side
of Eq.~(\ref{eq:2}) equals 2 when the modes of interest cross outside the horizon.

As a simple example, consider the case where $|V(Y)|\ll |E|$. 
Eq.~(\ref{eq:2}) then generically gives $(\ddot{a}_{pert}/a_{pert})\t^2\ll 1$.
This leads to a density spectrum of the form $|\delta_k|\sim k$, 
which is thus unacceptably blue. 
(A similar calculation was repeated for other set-ups 
such as Randall-Sundrum and the solutions presented in Ref.~\ref{kachru}. 
It was found that none of these solutions predict a scale-invariant spectrum of 
fluctuations when $V$ is turned off.) 
It is therefore crucial to add a potential in order to obtain a scale-invariant spectrum of density fluctuations.

\subsubsection{A successful example:  The exponential potential} \label{expon}

We can add a potential $V(Y)$ of 
the form that might result from the exchange of wrapped M2-branes.
We would like to think of $V$ as the potential derived from the superpotential $W$ for the 
modulus $Y$ in the 4d low energy theory. 
Typically, superpotentials for such moduli are of exponential form, for
example, 
\begin{equation}
W\sim e^{-cY},
\label{eq:W}
\end{equation}
where $c$ is a positive parameter with dimension of mass.
The corresponding potential is constructed from $W$ and the K\"ahler potential $K$ 
according to the usual prescription
\begin{equation}
V=e^{K/M_{pl}^{2}}\left[K^{ij}D_i W \,  \bar{D_j W}-\frac{3}{M_{pl}^{2}} 
W\bar{W}\right].
\label{eq:V}
\end{equation}
where $D_i = \partial/\partial \phi^i + K_i/M_{pl}^{2}$ is the K\"ahler covariant derivative,
$K_i = \partial K/ \partial \phi^i$, 
$K_{ij} = \partial^2 K/ \partial\phi^i \partial\phi^j$ and a sum over each
superfield $\phi_i$ is implicit.
Eqs.~(\ref{eq:W}) and~(\ref{eq:V}) imply that $V$ decays exponentially with 
$Y$.
For the purpose of this paper, we shall not worry 
about the exact form of the superpotential in heterotic M-theory. 
Rather, it will suffice to perform the calculation using 
a simple exponential potential, namely
\begin{equation}
V(Y)=-ve^{-\m\alpha Y},
\label{eq:expo}
\end{equation}
where $v$ and $\m$ are positive, dimensionless constants.
In this paper, it is convenient to parameterize the exponent in terms of $\alpha$.
Note that, in the case where the potential is generated by the exchange of wrapped
M2-branes, the parameter $m$ is of the form $m=c T_3\nu/\alpha$, where $c$ is a constant, 
$T_3$ is the tension of the M2-brane, and $\nu$ is the volume of the curve on which it is wrapped.
The potential defined in Eq.~(\ref{eq:expo}) is shown in Figure~\ref{fig:V}.
The perturbation modes of interest are those which are within the current
Hubble horizon. As the wavelengths corresponding to those modes passed
outside the effective Hubble horizon on the moving bulk-brane, the
amplitudes became fixed.  
Scale invariance will require $\m\D\gg 1$ during this period.
(In section~\ref{general}, we shall generalize this 
condition for an arbitrary potential $V$.)

We have already seen at the end of Section~\ref{scaleinv0} that, if the potential
$V$ is negligible compared to $E$, the spectrum of fluctuations is not scale-invariant.
Hence, we consider the limit where
 $|E| \ll | V_0| $.  This condition, as seen
from the equation of motion for $Y_0$, Eq.~(\ref{eq:back}), is satisfied if 
$\dot{Y}_0 =0$ initially, or, equivalently, if the bulk brane begins nearly 
at rest.  For the brane to be nearly at rest, one must have $|E| 
\approx |V_0|$ initially.  As the brane traverses the fifth dimension, $|V|$ increases 
exponentially, whereas $E$ is constant.  Hence, the condition $|E| \ll | V_0| $ is 
automatically satisfied.  The bulk brane beginning nearly at rest is precisely 
what we expect for a nearly BPS initial state.

Applying the condition  $|E| \ll | V_0| $, Eq.~(\ref{eq:t}) reduces to
\begin{equation}
\t^2\approx \frac{1}{2v}\left[\int_{0}^{Y_0}\D(Y')e^{\m\alpha Y'/2}dY'\right]^2\approx\frac{2\D_0^2}{\m^2\alpha^2ve^{-\m\alpha Y_0}}\left(1-\frac{2}{\m \D_0}\right),
\label{eq:t2}
\end{equation}
where we have neglected the endpoint contribution at $Y=0$.
On the other hand, Eq.~(\ref{eq:aeff}) gives
\begin{equation}
\frac{\ddot{a}_{pert}}{a_{pert}}=\frac{\m^2\alpha^2ve^{-\m\alpha Y_0}}{\D_0^2}\left(1+\frac{1}{\m\D_0}\right).
\end{equation}
Combining the above two expressions yields
\begin{equation}
\frac{\ddot{a}_{pert}}{a_{pert}}\t^2=2\left(1+\frac{1}{\m\D_0}\right)\left(1-\frac{2}{\m\D_0}\right).
\label{eq:2b}
\end{equation}
The right hand side
of Eq.~(\ref{eq:2b}) is approximately equal to 2 in the limit of large $\m\D_0$. 
Hence, the exponential potential of Eq.~(\ref{eq:expo}) results in a nearly
scale-invariant spectrum of perturbations provided that 
$|E| \ll |V_0|$ and $\m\D_0 \gg 1$ are satisfied 
when modes pass outside the effective Hubble horizon.
Note that it would be exceedingly 
difficult to maintain 
$(\ddot{a}_{pert}/a_{pert})\t^2$ close to almost any other 
value than 2.
 It is indeed fortunate that 
scale-invariance is the desired  result because  
obtaining a different spectral index from the ekpyrotic
scenario would be highly problematic.

We next compute the perturbation amplitude, by using Eq.~(\ref{eq:f1})
to calculate $|\Delta Y_k|$. 
As shown above, in order for the spectrum to be scale-invariant, the
conditions
$|E| \ll |V_0|$ and $\m \D_0 \gg 1$ must be  satisfied
when wavelengths pass outside the horizon.
These  conditions can be relaxed once the mode is well outside the horizon.
For example, we will assume no restrictions on $\m C$, the value of $\m\D_0$ at $y=0$.

In the limit that $\m\D_0\gg 1$ when the relevant modes cross outside the horizon, Eq.~(\ref{eq:f1}) reduces to 
\begin{equation}
x^2\frac{d^2f_{\vec{k}}}{dx^2}-\left[2-x^2\right]f_{\vec{k}}=0,
\label{eq:bessel}
\end{equation}
with solution
\begin{equation}
f_k=x^{1/2}\left(C_1(k)J_{3/2}(x)+C_2(k)J_{-3/2}(x)\right),
\end{equation}
where $J_{\pm 3/2}$ are Bessel functions.
The coefficients $C_1(k)$ and $C_2(k)$ are fixed by requiring that modes well-within the horizon 
(i.e., $x\gg 1$) be Minkowskian vacuum fluctuations, that is
\begin{equation}
f_{\vec{k}}=\frac{1}{\sqrt{6k\beta M_5^3B}}e^{-ik\t} \;\;\;{\rm for}\;\;x\gg 1.
\label{eq:cond}
\end{equation}
(Note that the factor of $\sqrt{6\beta M_5^3B}$ in Eq.~(\ref{eq:cond}) arises when we change 
variables from $Y$ to a canonically-normalized scalar field.)
Using this initial condition, we find the following rms amplitude for modes 
outside the horizon (with $x\ll 1$) 
\begin{equation}
\Delta f_k\equiv \frac{k^{3/2}f_k}{(2\pi)^{3/2}}=\frac{-i}{(-\t)(2\pi)^{3/2}\sqrt{6\beta M_5^3B}}.
\end{equation}
Substituting Eq.~(\ref{eq:t2}) and using $f_k=D_0\delta Y_k$, we find
\begin{equation}
\Delta Y_k=\frac{\m\alpha}{2(2\pi)^{3/2}\sqrt{3\beta M_5^3B}}\frac{\sqrt{ve^{-\m\alpha Y_0}}}{\D_0^2}.
\end{equation}
Finally, we define the time-delay $\Delta \t(k)$ by
\begin{equation}
|\Delta\t(k)|=\left\vert\frac{\Delta Y_k}{\dot{Y_0}}\right\vert=\frac{\m^2\alpha}{16\pi^{3/2}\sqrt{3\beta M_5^3B}}\left(\frac{2}{\m\D_0}\right),
\label{eq:delt}
\end{equation}
where we have used the equation of motion for $Y_0$, Eq.~(\ref{eq:back}).
Note that the time-dependence of $\Delta\t(k)$ is mild, a necessary condition for the validity of the time-delay formalism.
The factor of $\m\D_0\equiv \m\D(Y_0(\tau))$ 
is to be evaluated at time $\t$ when a given mode crosses
outside the horizon during the motion of the bulk brane.
Let $\D_k$ denote the value of $\D_0$ at horizon crossing for mode $k$.
Since horizon crossing occurs when $x=1$, or, equivalently, when $(-\t)=k^{-1}$, Eq.~(\ref{eq:t2}) gives
\begin{equation}
\D_k\approx\frac{2}{\m}\log\left(\frac{\m^2\alpha}{2k}\sqrt{\frac{ve^{\m C}}{2}}\right).
\label{eq:Dk}
\end{equation}
Substituting Eqs.~(\ref{eq:delt}) and~(\ref{eq:Dk}) into Eq.~(\ref{eq:dpot1}), we find
\begin{equation}
|\delta_k|=\frac{\alpha^2\m^4\sqrt{2v}}{4\pi^{3/2}\sqrt{3\beta M_5^3B}
(\m C+2)}\left(\frac{2}{\m\D_k}\right).
\label{eq:scal}
\end{equation}

This expression for $|\delta_k|$ increases gradually with increasing $k$, 
corresponding to a spectrum tilted slightly towards the blue.
The blue tilt is due to the fact that, in this example,
$D$ is decreasing
as the brane moves.
That is, the spectral index, 
\begin{equation}
n_s\equiv 1+\frac{d\log|\delta_k|^2}{d\log k}\approx 1+\frac{4}{\m\D_k},
\end{equation}
exceeds unity.
The current CMB data constrains the spectral index to lie in the range 
about $0.8<n_s<1.2$.
Therefore, for our results to be consistent with experiments, we must have
\begin{equation}
\m\D_k>20,
\end{equation}
a constraint that is easily satisfied.

\subsubsection{General potential} \label{general}

As a second example, consider the power-law potential
\begin{equation}
V(Y)=-vD(Y)^q=-v(\alpha Y+C)^q,
\label{eq:power}
\end{equation}
where $v>0$ and $q<0$ are constants. In this case, Eq.~(\ref{eq:2}) gives
\begin{equation}
\frac{\ddot{a}_{pert}}{a_{pert}}\t^2\approx 2\frac{\left(1-\frac{2}{q}\right)}{\left(1-\frac{4}{q}\right)^2}\approx 2
\end{equation}
for $|q|\gg 1$.
Hence, a power-law potential can also lead to a nearly scale-invariant spectrum provided that its exponent is sufficiently large.
(The smaller is the value of $|q|$, the bluer is the spectrum.)

We can straightforwardly extend our analysis to an arbitrary potential $V(Y)$.
Let us suppose that $V(Y)$ satisfies
\begin{eqnarray}
\nonumber
& & \left\vert D(Y)\frac{dV}{dY}\right\vert\gg \alpha|V(Y)| \\
& & \left\vert D(Y)\frac{d^2V}{dY^2}\right\vert\gg \alpha \left\vert\frac{dV}{dY}\right\vert,
\label{eq:derivcond}
\end{eqnarray}
(For the exponential potential, $V(Y)=-ve^{-\m\alpha Y}$, these two conditions amount to $\m\D\gg 1$.)
Then, Eq.~(\ref{eq:2}) reduces to 
\begin{equation}
\frac{\ddot{a}_{pert}}{a_{pert}}\t^2\approx 2\left(\frac{VV''}{V'^2}\right).
\end{equation}
Hence, the conditions for scale invariance 
are Eqs.~(\ref{eq:derivcond}) as well as
\begin{equation}
\frac{VV''}{V'^2}\approx 1.
\end{equation}

\subsection{Gravitational waves from colliding branes}

In inflationary cosmology, the analysis of tensor (gravitational wave) 
perturbations follows closely the analysis of scalar (energy density) 
perturbations~\cite{Mukh2}. Metric fluctuations can be divided into two 
polarizations, each of which acts like a massless scalar field evolving
in the same cosmic background as the inflaton.
Hence, it is not surprising that the spectrum of tensor fluctuations has nearly 
the same scale-invariant spectral shape as the scalar spectrum.

In the ekpyrotic scenario, the relationship between scalar and tensor 
perturbations is less direct. The excitations that produce scalar perturbations 
are ripples on the moving brane, which are directly dependent on the rate at 
which the brane traverses the fifth dimension and the potential that drives it.
The tensor fluctuations, on the other hand, are excitations of the gravitational 
field, which lives in the bulk.  (The moving brane itself does not support 
tensor fluctuations.)  The net result is a different effective scale factor 
in the fluctuation equation of motion for the tensor modes than for the scalar modes.

We shall briefly outline the derivation here, 
with more details to follow in our 
more formal paper on  perturbations~\cite{pert}.
If $\bar{g}_{\mu \nu}$ is the unperturbed, homogeneous metric (see Eq.~(\ref{eq:static}) with $A$ and $N$ functions of time), the perturbed 5d 
metric can be written as
\begin{equation}
g_{\mu \nu} = \bar{g}_{\mu \nu} + A^2(t) D(y,t)h_{\mu \nu}(\vec{x},t),
\end{equation}
where $\mu,\nu=0,\dots 3$.
Note that since we shall work at the level of the 4d effective theory, we can treat the tensor perturbations $h_{\mu \nu}$ as functions of $\vec{x}$ and $t$ only.
We are interested in the tensor perturbations which satisfy the conditions:
$h_{0 \mu}=0$, $h^i_j=0$, and $\partial^i h_{ij}=0$. 
The perturbed 5d Einstein action to quadratic order is 
\begin{equation}
S_{fluct}^T \equiv \frac{M_5^3}{2}\int d^5 x \sqrt{-g} R =  \frac{M_5^2}{8}\int 
d^4 x \a^2 (\dot{h}^{\mu}_{\;\nu} \dot{h}^{\;\nu}_{\mu} - \partial_i  
h^{\mu}_{\;\nu} \partial^i  h^{\;\nu}_{\mu})
\end{equation}
where the second expression is obtained by integrating over $y$.  
The tensor 
action is analogous to the scalar action given in Eq.~(\ref{eq:flucaction}).
From the action, we can derive the tensor analogue of the scalar fluctuation 
equation of motion, Eq.~(\ref{eq:f1})
\begin{equation}
x^2 \frac{d^2 f_{\vec{k}}^T}{d x^2} - \left[ \frac{\ddot{\a}}{\a} \t^2 -
x^2\right] f_{\vec{k}}^T=0,
\end{equation}
where 
\begin{equation} 
h^{\mu}_{\;\nu} \equiv \int \frac{d^3 k}{(2 \pi)^3} \epsilon^{\mu}_{\;\nu} 
h_k (\t)
\end{equation}
and
\begin{equation}
f_{\vec{k}}^T \equiv \a \, h_{\vec{k}}.
\end{equation}
The critical difference between this tensor
equation and the scalar fluctuation equation,
Eq.~(\ref{eq:f1}), is that the 
effective scale factor $a_{pert}$ in Eq.~(\ref{eq:f1}) has been replaced by $\a$.
We introduced a potential to insure that $a_{pert}$ 
led to a nearly scale-invariant spectrum, $(\ddot{a}_{pert}/a_{pert})\t^2\approx 2$.
However, $\a(\t)$ in the tensor equation is approximately constant (recall that $\a=(BI_3^{(0)}M_5)^{1/2}+{\cal O}(\beta/\alpha)$).
Consequently, the root mean square tensor fluctuation amplitude,
\begin{equation}
|\Delta h_{\vec{k}}| \equiv \frac{k^{3/2} h_{\vec{k}}}{(2 \pi)^{3/2}} \sim 
\frac{k}{(2\pi)^{3/2}}.
\end{equation}
is not  scale-invariant. Rather, the tensor spectrum is tilted strongly 
to the blue.  Fitting the mean square amplitude to a scale-free form, $\sim 
k^{n_T}$, where $n_T$ is the conventional tensor spectral index, the spectrum 
above corresponds to $n_T=2$, compared to the inflationary prediction, $n_T \le 
0$.

The tensor spectrum is a prediction that clearly distinguishes the ekpyrotic
 scenario from inflationary cosmology.  In both cases, for the same Hubble 
parameter at reheating (for inflation) or collision, $\H_{c}$, the mode with 
wavelength of order $\H_{c}^{-1}$ has similar amplitude, $\H_{c}/M_{pl}$,
where $M_{pl}$ is the 4d Planck mass.  
The wavelength of this mode today
is roughly 60 e-folds smaller than the current Hubble radius.
Hence, if we extrapolate from this wavelength to one comparable to the present 
Hubble radius, $H_0^{-1}$, the inflationary prediction is that the amplitude is 
nearly the same (since the spectrum is nearly scale-invariant), whereas the blue 
spectrum computed above predicts that the amplitude is exponentially small.  
Hence, the search for a gravitational wave signal using the CMB polarization on 
horizon scales becomes a key test for our proposal.  
Future gravitational wave detectors, beyond the presently planned LISA and LIGO projects, 
may also someday detect the stochastic background of gravitational waves as well.
Observing a nearly scale invariant primordial gravitational 
wave background falsifies the 
ekpyrotic scenario and is consistent with inflation.  

\section{Conclusions}

\subsection{Recapitulation}

Conceptually, the ekpyrotic scenario appears to be simple:  a bulk 
brane strikes our visible brane and a hot big bang universe is born.
In actuality, to accomplish the transformation 
from a cold, nearly BPS state into an expanding, hot universe
with nearly scale-invariant density perturbations without
invoking inflation requires a series of seemingly incongruous
conditions. Remarkably, these conditions can be satisfied simultaneously in heterotic M-theory.

First, the gravitational backreaction due to the kinetic energy 
of the bulk brane must trigger  cosmic expansion.
In 4d gravity, kinetic energy usually causes cosmic deceleration
and, if the initial state is static, it triggers contraction.
Second, the total energy of the brane has to grow by drawing
energy from the gravitational field since, otherwise, the total
bulk brane
energy before and after collision is zero and there is no radiation.
 In 4d gravity, this blue
shift effect occurs if the universe is contracting, but here
it occurs even though the scale factor on the visible brane is expanding.
Third,  the 
scalar (energy density) fluctuations must be nearly scale-invariant.
Although scale-invariance is ordinarily associated with inflation, here
we have shown that a scale-invariant spectrum results even
though the universe is quasi-static.  
The only requirement is a bulk brane potential
whose magnitude increases by an 
exponential factor as the brane traverses the bulk. 
Potentials of this type occur in string theory.
(Our analysis suggests that scale-invariance is especially
favored in the 5d theory. It  occurs for
rather simple,  physically-motivated potentials, whereas
more general spectral shapes are difficult to obtain.)
Fourth, the tensor fluctuation spectrum is
not scale-invariant but, rather, strongly tilted towards
the blue, providing an  observational signature
that distinguishes the ekpyrotic scenario
from inflationary cosmology.

In the context of 4d gravity,
some of these features suggest slow 
expansion, others superluminal expansion, yet others 
contraction. How do we obtain all of these features simultaneously?
All of this is possible in our 5d theory
because  the role of gravity in
the equations of
motion  for the expansion, the brane motion, and the scalar and
tensor fluctuations  is played  by different combinations of moduli fields
 for each equation.
That is, each equation is similar
to the corresponding equation for  a scalar field in a 4d FRW
background except that the scale factor
$a(t)$ is replaced by   some function of
the   moduli fields that differs for each physical quantity.
Some combinations increase
with time, mimicking an expanding universe,  and others decrease,
mimicking a contracting universe. 
The remarkable result is that just the  right  combinations of moduli
 occur  in heterotic M-theory  to produce 
 the  behavior required for 
 a viable cosmological scenario.

An observer at any surface of fixed $y$ has a scale factor equal
to $AD^{1/2}(y,\t)$ (see Eq.~(\ref{eq:static})), which is expanding
as the bulk brane collides with the visible brane.
In Eq.~(\ref{eq:psi}), which describes the time variation of the
total energy of the bulk brane, the role of the scale factor is
played by $\a$, which is contracting.
The contraction produces the blue shift or increase in the total
energy so that, upon collision, there is excess kinetic energy 
that can be converted to radiation.
The scalar fluctuations are ripples in the bulk brane surface that
evolve as if the scale factor were $a_{pert}$ in Eq.~(\ref{eq:f1}),
corresponding to a contracting effective Hubble radius.
We have identified simple criteria for the
potential which result in a scale-invariant spectrum.
The tensor fluctuation spectrum naturally differs from the scalar spectrum
because tensor fluctuations occur in the bulk volume rather than on 
the brane surface. The effective scale factor for the tensor 
fluctuations is $\a$, rather than $a_{pert}$. 
The differences account for the fact that 
the scalar spectrum is scale-invariant, whereas the tensor 
spectrum is tilted strongly to the blue.

A useful mnemonic for recalling the difference between the scalar
and tensor fluctuation spectra in our scenario
is to consider the equivalent relations for
inflationary cosmology, but with the inflaton scalar field replaced
by $Y$.  For the scalar fluctuations, the amplitude is
\begin{equation}
\delta_S \sim H\left(\frac{\Delta Y}{\dot{Y}}\right)_{k=H}
\end{equation}
and the tensor fluctuation amplitude is
\begin{equation}
\delta_T \sim \left(\frac{H}{M_{pl}}\right)_{k=H},
\end{equation}
where  $M_{pl}$ is the 4d Planck mass, $\Delta Y$ is
the fluctuation amplitude for  $Y$, and the subscript means
that the expressions are to be evaluated
when the wavenumber of a given
mode is equal to the inverse Hubble radius $H$ as it passes beyond
the horizon.
For inflation, $H$, $\Delta Y$ and $\dot{Y} $
are nearly constant, so $\delta_S$ and
$\delta_T$ are both nearly scale-invariant. For the ekpyrotic
scenario, $ \Delta Y$ and $\dot{Y}$ are both strongly time-varying.
However, for an exponential potential $V(Y)$  as is naturally
generated by non-perturbative exchange of M2-branes,
the time-variation
in the ratio $\Delta Y/\dot{Y}$ in the expression for $\delta_S$
nearly cancels.
Consequently, the ratio
is nearly constant and the resulting spectrum is nearly scale-invariant.
However, $\delta_T$ involves only $H$, which is increasing with time
as smaller and smaller wavelength modes pass beyond the horizon.
This accounts for the fact that the spectrum is blue.

As a specific fully-worked example, 
consider the case of an exponential bulk brane
potential, $V(Y)=-v \, {\rm exp}(- \m \alpha Y)$, as discussed
in Section~\ref{expon}. We have computed the ekpyrotic temperature
at the beginning of the hot big bang phase (Eq.~(\ref{eq:Texpo})),
\begin{equation}
\frac{T}{M_{pl}}\approx
\frac{3^{3/4}(2v)^{1/4}}{(I_3M_5)^{1/2}(\alpha R+C)^{1/4}}\left(\frac{M_5}
{M_{pl}}\right)^{1/2}\left(\frac{\beta}{\alpha}\right)^{1/2}\frac{(\m C+2)^{1/2}}{m}.
\end{equation}
In terms of this temperature, the scalar (energy density) fluctuation
amplitude in Eq.~(\ref{eq:scal}) can be rewritten as
\begin{equation}
|\delta_k|=\frac{\m^6 (I_3\alpha)^{3/2}}{36\pi^{3/2} (\m C +
2)^2}\left(\frac{\alpha}{\beta} \right)^{3/2}
\frac{2}{\m\D_k}\left(\frac{T}{M_{pl}}\right)^2.
\end{equation}

A simple example which satisfies all
constraints is $\alpha=2000M_5$, $\beta=M_5$, $B=10^{-4}$, 
$C=1000$, $R=M_5^{-1}$, $\m=0.1$, and $v=10^{-8}$,
all of which are plausible values. In this example, $D_k$ (the value
of $D$ at horizon crossing) is of order $10^3$.
Then, we find that 
$M_5 \sim 10^{-2} \, M_{pl}$; 
the ekpyrotic temperature is $T\sim 10^{-8} \, M_{pl}$; and
the scalar perturbation amplitude is 
$|\delta_k| \sim 10^{-5}$.  Note that the ekpyrotic temperature,
the maximal temperature of the hot big bang phase, tends to be small
compared to the Planck or unification scale.
This is a characteristic feature of the model.
With these parameters, the magnitude of the potential energy density
at collision is $(10^{-6}M_{pl})^4$.
This corresponds to a characteristic energy scale for the potential of $10^{13}$~GeV.
Finally, note that these values are consistent with Ho\v rava-Witten phenomenology.
For instance, the proper distance between the branes is $R_{proper}^{-1}\sim 10^{-5}M_{pl}$.
If we further assume that the characteristic length scale $L_{CY}$ of the 
Calabi-Yau three-fold is approximately 10 times smaller than $R_{proper}$ (in order for the five-dimensional effective theory to be valid), then we get agreement with the values of $R_{proper}$, $L_{CY}$, and the 11d Planck mass $M_{11}$ inferred by Witten in Ref.~\ref{witten2} in matching the gauge and gravitational coupling constants.

While we are pleased that the numerical constraints from cosmology 
and those from Ho\v rava-Witten phenomenology can be 
simultaneously satisfied, we should emphasize that there is 
a lot of flexibility in terms of parameters. 
For instance, in the above example the ratio $\beta/\alpha$ was chosen to be of order $10^{-3}$.
However, one can easily make this ratio as large as $1/10$ if one wishes.
For example, choosing $\alpha=200M_5$, $\beta=20M_5$, $B=10^{-2}$, 
$C=100$, $R=M_5^{-1}$, $\m=1$, and $v=10^{-10}$ results in $T\sim 10^{-7} \, M_{pl}$,
$|\delta_k| \sim 10^{-5}$, $M_5 \sim 10^{-2} \, M_{pl}$, $R_{proper}^{-1}\sim 10^{-5}M_{pl}$, and
characteristic energy scale for the potential of $10^{14}$~GeV.
The value of $R_{proper}$ could also take a significantly different value, if one wishes, and 
there would still be enough freedom to obtain reasonable
ekpyrotic temperature and fluctuations. We can also imagine applying 
the same ideas in a different brane world context, such as AdS, and still obtaining a 
successful scenario from a cosmological point-of-view. The challenge, 
of course, is to figure out how to break supersymmetry, obtain a 
correct phenomenology and stabilize moduli. Here we have presumed that
this challenge can be met, 
and have shown through examples how the brane world approach to 
particle phenomenology might be 
combined with  new ideas in cosmology to obtain  a successful picture of the early Universe.

\subsection{Colliding Branes and Inflationary Cosmology}

We have laid out a detailed cosmological
scenario that offers a resolution of the flatness, horizon, and monopole
problems and generates a nearly scale-invariant spectrum of energy density 
perturbations based on concepts that derive naturally from extra dimensions, 
branes,
and heterotic M-theory. 
The key conceptual difference from inflation is how the universe 
begins. In the usual approaches to 
inflationary cosmology, as in standard big bang cosmology,
the universe begins with a cosmological singularity. The universe
emerges in a high energy state with no 
particular symmetry and rapidly expanding.  Superluminal expansion
is invoked to smooth out and flatten  the emerging state.
The ekpyrotic scenario introduces a different philosophy in which 
the universe begins in a non-singular,
infinite, empty, quasi-static state of high 
symmetry.  
Superluminal expansion is not needed because
the BPS vacuum state is  flat and smooth.  
Brane collision  can account for 
the matter-radiation energy and primordial 
density perturbations.

Let us briefly 
summarize how the ekpyrotic  scenario addresses the
various cosmological
problems: 

\begin{itemize}
\item {\it Causal Horizon Problem:}
In the ekpyrotic scenario,
the local temperature and density are set by the  collision
of the visible brane and
 bulk brane,
which  acts as a non-local event that occurs
nearly simultaneously over a region much larger than the Hubble horizon.
\item{\it Flatness Problem:}  The universe is assumed to begin in a
nearly BPS ground state.  The BPS state 
corresponds to a spatially-flat geometry.
The process of bulk brane formation/nucleation and propagation maintains
flatness. 
(We do not demand that the initial state be globally BPS to resolve
the horizon and flatness problems.
It suffices that the universe be flat and homogeneous  on
scales up to  the (causal) particle horizon, as should
occur naturally beginning from more general initial conditions.
In the ekpyrotic scenario, because the bulk brane motion is 
extremely slow, the particle horizon at collision is
exponentially large compared to the Hubble horizon,
where the latter is set by the radiation temperature after collision.)
\item{\it Monopole Problem:} The hot big bang epoch commences when the bulk
brane collides with the
visible brane and heats the universe to a finite 
temperature. Provided the temperature is less than the monopole mass, the
monopole abundance will be negligible.
\item {\it Inhomogeneity Problem:}  
Quantum fluctuations
generate ripples in the bulk brane as it traverses the bulk.
 Due to the ripples, collision and thermalization occur at varying times
across the visible brane, resulting in 
fluctuations in energy density and gravitational waves.
\end{itemize}

Both the ekpyrotic scenario and inflationary cosmology have
the feature that the  causal horizon 
is exponentially  greater than the Hubble
horizon.  In inflation,
superluminal expansion rapidly stretches the 
causal horizon while the  Hubble horizon is nearly fixed.   
In the colliding brane picture, the collision 
of the bulk brane acts as a non-local interaction that causally links
regions separated by much more than a Hubble distance.

Both the inflationary and ekpyrotic scenarios produce 
a nearly scale-invariant spectrum of energy density perturbations from 
quantum fluctuations.  For inflation, quantum fluctuations are stretched
beyond the Hubble horizon as the universe expands superluminally. 
For the ekpyrotic universe, the Hubble horizon is
shrinking compared to the quantum fluctuations as the universe contracts
very slowly.  The equations describing the evolution of 
perturbations are (nearly)  equivalent in 
the two cases (see discussion of Eq.~(\ref{eq:f1})), even though, 
one describes an expanding de Sitter phase and the other a contracting
pressureless phase.  The similar equations account for why both 
lead to scale-invariant spectra for density perturbations
even though the mechanisms are different.

From the point-of-view of an observer on the stationary orbifold planes, the 
universe is expanding  as the branes collide.
 The bulk brane is what causes their expansion, a gravitational
backreaction effect due to its motion.
The expansion  is very slow as the brane moves across the fifth 
dimension, but assumes the usual big bang rate after collision and
thermalization.

One might hope that the ekpyrotic scenario avoids 
the tuning problems required in standard inflation in order to obtain
an acceptable perturbation spectrum.  Thus far, the situation
is unclear.  We found that we had to introduce a flat potential
for the bulk brane that is roughly similar to the flat inflaton potential 
used in standard inflation. 
The form is also qualitatively consistent with non-perturbative 
potentials that arise in M-theory. 
Perhaps the potential parameters needed for our scenario
will be shown to arise naturally.
However, it should also be noted that  the reasons for introducing the 
potential in the ekpyrotic scenario are different from the case of inflation.
In our case, the need for a flat potential is linked to the 
precise form of the 
background static BPS solution of Lukas, Ovrut and Waldram used in this
paper.  
Perhaps  there  exist other initial conditions which avoid the need for 
flat potentials altogether.

Although inflationary cosmology and
the colliding brane picture  both 
produce a nearly scale-invariant spectrum of perturbations, the deviation
from scale-invariance differs due to  the fact that the background felt by the perturbations is 
expanding  in one scenario and contracting in the other.   
In standard inflationary cosmology, 
the spectrum of scalar (density) and tensor (gravitational wave) perturbations
is typically red (amplitude
decreases as wavelength decreases).\cite{lidsey}
The amplitude of a
given mode is proportional
to the Hubble parameter when the  wavelength is stretched beyond the 
horizon.  The Hubble parameter  decreases (slowly)  in an expanding, inflating 
universe. Since smaller wavelength modes stretch beyond the horizon
at later times when the Hubble parameter is smaller, they have a 
smaller amplitude, resulting in a red spectrum.
The degree of redness is expressed in terms of a ``spectral index,'\cite{Davis}
$n_{S,T}$ for scalar and tensor perturbations respectively, where
$n_S-1 =n_T =0$ is defined as precise scale-invariance, and 
$n_S  -1 <0$ and $n_T<0$ correspond to red spectra. 
In the examples of the ekpyrotic scenario discussed here, 
the apparent  Hubble radius 
for an observer on the bulk brane is shrinking.
Consequently, the corresponding spectra are blue.
By introducing a potential for the bulk brane
(dependent, say, on its position $Y$), the density perturbation spectrum
can be made nearly scale-invariant, 
slightly blue
($n_S-1 >0$) in our  examples. 
On the other hand, the gravitational wave spectrum
is unaffected by the potential and is strongly blue $n_T\approx 2$.

For energy density perturbations, there are exceptional cases where
inflation can  give a blue spectrum~\cite{hybrid}.
The blue spectrum arises because
the density perturbation amplitude is not only proportional to the Hubble
parameter, but also inversely proportional to the 
kinetic energy of the inflaton.
As inflation proceeds,
the Hubble parameter decreases and, in most models, the inflaton 
kinetic energy increases; so both effects 
tend to make the spectrum red.  But models can be rigged where the
Hubble parameter decreases, as usual, but the 
inflaton kinetic energy decreases more rapidly.
In that case, the spectrum is blue.
Similarly, it is possible to get a red spectrum in the ekpyrotic 
model, for example, if the bulk brane moves in the direction of 
increasing warp factor.\cite{KKL}
Hence, 
observing a red or
blue density spectrum is not a decisive test for distinguishing
the two scenarios.

However, the gravitational wave spectrum for 
inflation is always red  --- the amplitude depends only on the Hubble
parameter --- and so observing a strongly blue gravitational wave spectrum,
as predicted by the ekpyrotic scenario, is 
a key test.   The cosmic microwave background polarization is 
one method of detecting the presence of primordial gravitational waves
with wavelengths comparable to the Hubble horizon today.  For the slightly
red spectrum of inflationary cosmology, 
the gravitational wave amplitude may be large enough to 
be detected. However, for a strongly blue spectrum, the signal at large length
scales is exponentially small and  undetectable. Hence, the detection of
gravitational waves in the microwave background polarization would falsify
our scenario.
Currently planned  gravitational wave detectors, such as LIGO and 
LISA, are not sensitive enough  to detect the gravitational 
waves from inflation or from our scenario. 
However, future detectors may discover the stochastic background and 
determine the spectral slope.

Certain aspects of our scenario are reminiscent of  the novel,
string-inspired, pre-big bang models introduced by 
Veneziano and Gasperini~\cite{Venez,Gasp}. 
Both assume the universe begins in  a flat, empty state~\cite{APT}.
In both models, 
 the gravitational wave perturbation spectrum is blue~\cite{bru}.
However, the structure, ingredients,  dynamics and predictions 
of the two models
are very different.
The pre-big bang scenario
does not entail extra dimensions or branes in
a direct way. The pre-big bang begins with a semi-infinite period
of contraction which is superluminal (deflation).  This period
ends in a global singularity in which the Hubble constant 
becomes infinite. Matching across this singularity is 
the biggest challenge facing the pre-big bang model.
If it is possible at all, it can only be at the string scale
where non-perturbative stringy
effects are large and difficult to compute.
In our model, matching to the universe after the bulk-boundary
collision is complex, and will require calculations in 
the five dimensional theory which we have not studied here.
But the collision event itself is nonsingular and it
is an important advantage of our scenario that it
only involves physics taking place at low energies,
which is in principle describable using the
effective low energy Lagrangian. 
A second difference is that all
expansion and contraction are subluminal in our model.
Finally, in the simplest renditions of pre-big bang,
the fluctuation spectrum is so strongly tilted to the blue relative to 
a scale invariant spectrum that current observational
bounds on the microwave background anisotropy are violated.
The ekpyrotic scenario obtains a spectrum that is consistent with 
current observations.

 To summarize, we have presented a novel scenario for the beginning of the 
 hot big bang 
 universe, within a framework consistent with string theory and supergravity.
The universe begins in the simplest state possible, one which is cold, nearly 
BPS, and nearly vacuous.
At some time, a bulk brane exists or is nucleated in the vicinity of the hidden
brane
(through a small instanton phase transition), and begins to move towards the 
visible brane.
The bulk brane eventually collides with the visible brane and is absorbed in a 
small instanton phase transition. 
This transition may change the gauge group on the visible brane to the standard
model
gauge group, as well as create three families of light quarks and 
leptons.
At the moment of collision, a fraction of the kinetic energy of the brane is 
converted to thermal excitations of the light degrees of freedom on the visible 
brane, causing the universe to enter an FRW radiation-dominated phase.
Furthermore, ripples on the bulk brane imprint a spectrum of energy density 
fluctuations consistent with current observations and which provides the seeds 
for structure formation.
While parts of our scenario remain speculative at present (such as the dynamics 
of the small instanton phase transition), it is our hope that advances in 
heterotic M-theory will eventually allow us to solidify the 
components of our cosmological model.
For the moment, we consider our scenario as a first step 
towards a new, testable 
model for the early universe consistent 
with current cosmological observations and fully-motivated by string theory.

\bigskip
\noindent 
{\sc Acknowledgments} \\
We would like to thank C.P. Burgess, S. Gubser,
 J. Maldacena, M. Bucher, N. Seiberg, D. Spergel,
  D. Waldram and T. Wiseman for helpful discussions, and
   Katharina Volk and J. Katz for introducing us to
    ancient cosmology and ekpyrosis.
We thank R. Kallosh {\it et al.}~\cite{KKL,kallosh} for pointing out minor
typographical errors in the manuscript.
    We would also like to thank the Director and staff
    of the Newton Institute for Mathematical Sciences, where this work was
    initiated, and the organizers of the String Cosmology Workshop at the
    the Pacific Institute for Mathematical Sciences (Vancouver) where
    some of this work was carried forth.
    This work was supported in part by
     the Natural Sciences and Engineering Research Council of
     Canada (JK),
     the  US Department of Energy grants
     DE-FG02-91ER40671 (JK and PJS) and DE-AC02-76-03071 (BAO), and
     by PPARC-UK (NT).

\pagebreak

\end{document}